\documentclass[lettersize,journal]{IEEEtran}
\usepackage{amsmath,amsfonts}
\usepackage{algorithmic}
\usepackage{algorithm}
\usepackage{array}
\usepackage{textcomp}
\usepackage{stfloats}
\usepackage{url}
\usepackage{verbatim}
\usepackage{graphicx}
\usepackage{soul}
\usepackage{cite}
\DeclareMathAlphabet\mathbfcal{OMS}{cmsy}{b}{n}
\usepackage{algorithm}      
\usepackage{algorithmic}

\usepackage{varwidth}
\usepackage{float}
\usepackage{graphicx}
\usepackage{subcaption}
\usepackage{balance}
\usepackage{xcolor}
\usepackage{booktabs}
\usepackage{bbm}
\usepackage{array, multirow}
\usepackage{adjustbox}
\usepackage{url}
\usepackage{hyperref}
\def\etal{\emph{~et~al.}}
\def\BibTeX{{\rm B\kern-.05em{\sc i\kern-.025em b}\kern-.08em
    T\kern-.1667em\lower.7ex\hbox{E}\kern-.125emX}}
    
\usepackage{textcomp}
\usepackage{amsmath,amssymb,amsfonts}

\begin{document}
\title{
Augmented Driver Behavior Models for High-Fidelity Simulation Study of Crash Detection Algorithms
}
\author{Ahura Jami$^{1}$, Mahdi Razzaghpour$^{1}$, Hussein Alnuweiri$^{2}$, Yaser P. Fallah$^{1}$
\thanks{$^{1}$ Connected \& Autonomous Vehicle Research Lab (CAVREL), University of Central Florida, Orlando, FL, USA.}
\thanks{$^{2}$ Department of Electrical and Computer Engineering, Texas A\&M University at Qatar, Doha, Qatar}
\thanks{This research was supported in part by the Qatar National Research Fund Project NPRP 8-1531-2-651.}
}

\maketitle
\begin{abstract}
Developing safety and efficiency applications for Connected and Automated Vehicles (CAVs) requires a great deal of testing and evaluation. The need for the operation of these systems in critical and dangerous situations makes the burden of their evaluation very costly, possibly dangerous, and time-consuming. As an alternative, researchers attempt to study and evaluate their algorithms and designs using simulation platforms. Modeling the behavior of drivers or human operators in CAVs or other vehicles interacting with them is one of the main challenges of such simulations. While developing a perfect model for human behavior is a challenging task and an open problem, we present a significant augmentation of the current models used in simulators for driver behavior. In this paper, we present a simulation framework for a hybrid transportation system that includes both human-driven and automated vehicles. In addition, we decompose the human driving task and offer a modular approach to simulating a large-scale traffic scenario, allowing for a thorough investigation of automated and active safety systems. Such representation through Interconnected modules offers a human-interpretable system that can be tuned to represent different classes of drivers. Additionally, we analyze a large driving dataset to extract expressive parameters that would best describe different driving characteristics. Finally, we recreate a similarly dense traffic scenario within our simulator and conduct a thorough analysis of various human-specific and system-specific factors, studying their effect on traffic network performance and safety.
\end{abstract}


\section{Introduction}
\noindent Recent studies of traffic data collected on US roadways have shown a steady increase in road fatalities, resulting in over 35,000 lives lost and over \$240 billion in damages each year\cite{Sauber2016}. Although motor vehicle accidents constitute one of the highest causes of death in the USA\cite{Webb2012}, they are almost always predictable and preventable. Deploying Advanced Driver Assistant Systems (ADAS) within an intelligent transportation framework is a low-cost, yet effective, step toward resolving this issue. Depending on the level of autonomy of such safety systems, they can either warn drivers by recognizing potentially dangerous situations or take over vehicle control through prevention mechanisms. Forward collision and lane departure warnings are two examples of active safety systems that provide advisory alerts. Forward collision avoidance and lane-keeping assistance are more advanced versions of these systems that provide automated control in addition to recognizing dangerous scenarios.

Vehicle instantaneous position, kinematic information (heading, velocity, acceleration, yaw rate, etc.), driver-specific information (head pose, eye gaze, etc.), and environmental data (road type, weather conditions, lighting characteristics, etc.) are some of the most informative situational elements used by ADAS\cite{Sengupta2007, Malaterre1998}. 
In general, the data used by an ADAS can be separated into two groups: the host vehicle's data and its surrounding objects. The host vehicle's positional and kinematic data can be collected from the Global Positioning System (GPS) and the vehicle's Controller Area Network (CAN) bus, which provides sensory information such as speed, acceleration, throttle position, and steering angle. The second type of data, namely information about remote objects, is obtained using local sensors such as cameras, radar, and LiDAR, as well as wireless communication between vehicles, known as Vehicle-to-Vehicle (V2V) communication. The current version of ADAS, which is already in use in some high-end vehicles, scans the environment and monitors driving conditions using local sensors. Recent communication advancements in wireless vehicular networking offer an opportunity for the automotive industry to be on the brink of a breakthrough in automation and safety improvements through designing and utilizing Cooperative Vehicular Safety (CVS) applications. As a cyber-physical system, a modern vehicle uses V2V communication as a desirable platform for developing a rich, advanced, and low-cost safety system\cite{Sengupta2007,fallah2012cyber, Moradi2017, Tahmasbi2017}.
Furthermore, using Vehicle-to-Everything (V2X) communication will expand vehicles' perception beyond the range of their onboard sensors.

However, integrating such automated mechanisms into a traditional transportation system poses several technical challenges. Given the stochastic nature of host or remote vehicle drivers, it is critical for automated systems to have a detailed awareness of their surroundings. Furthermore, in order to improve the performance and efficiency of such systems, they must be designed within a flexible framework that takes driver behavior models into account. Another crucial issue that overshadows the design of an adaptive system is the need for rich and extensive driving data in which these systems are to perform. When approached through real-world measurement campaigns, the data collection phase can be very costly and time-consuming\cite {9334665, SPMD:Data}. Furthermore, some of the most critical test scenarios can be extremely dangerous because they expose real drivers to potentially hazardous situations. A thorough analysis of driver behavior during a collision scenario, for example, is required to design an adaptive Forward Collision Warning (FCW) algorithm. A potentially dangerous collision scene should be designed and tested in a series of trials. These critical issues motivate the development of a versatile simulation platform capable of performing any examination without the use of real drivers.

\begin{figure*}
    \centering
    \resizebox{14cm}{!}{\includegraphics[]{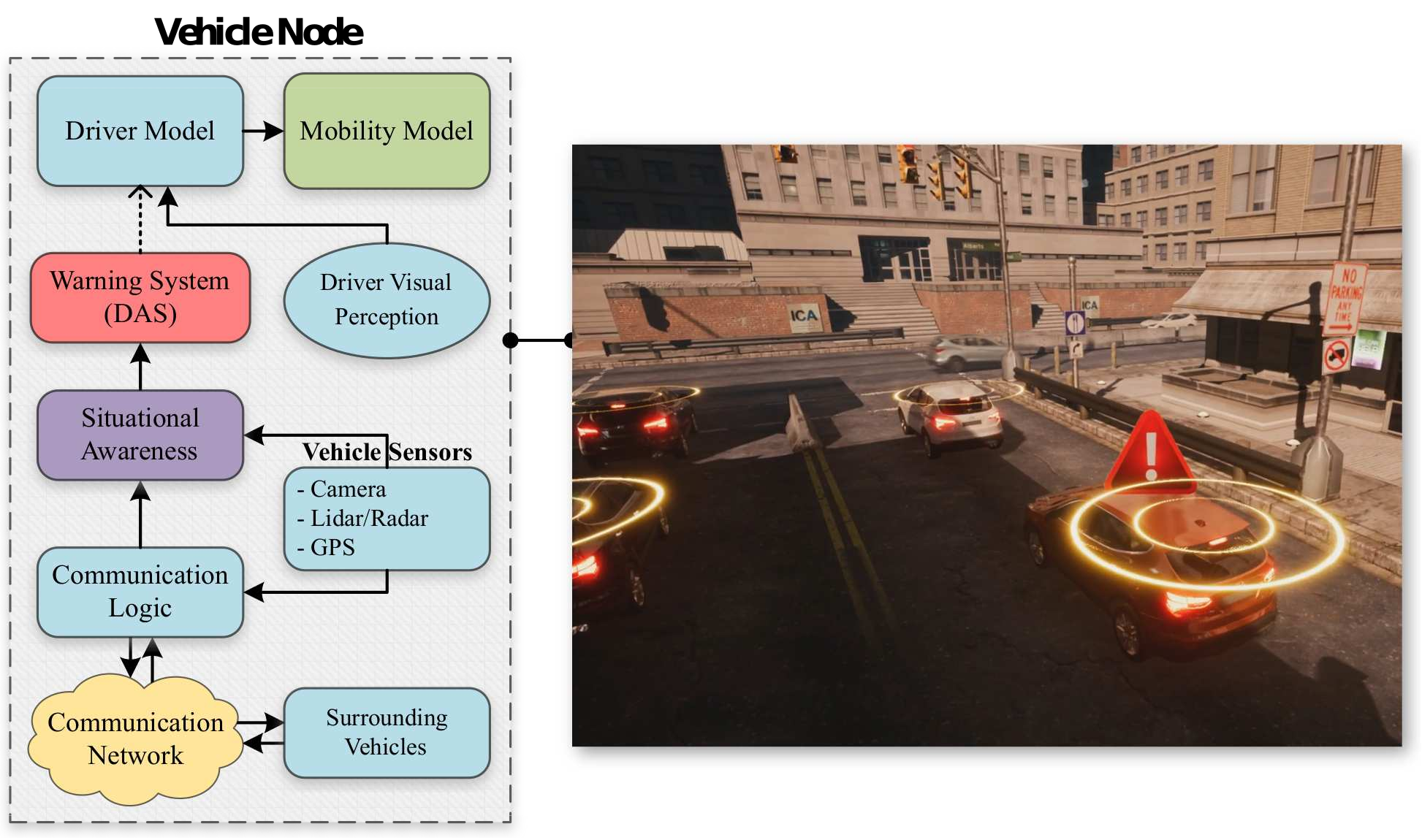}}
    \caption{An architectural overview of the co-simulation framework along with a snapshot of the simulator environment}
    \label{fig:architecture_overview}
\end{figure*}

Despite the fact that an Intelligent Transportation System (ITS) is multidisciplinary, traditional simulators were designed solely for studying traffic management topics. They have limited capabilities for assessing active safety mechanisms in life-threatening situations. The analysis of such critical systems necessitates a comprehensive framework that allows for an assessment with sub-second accuracy in various traffic conditions.
There have been several attempts to incorporate novel ITS applications into traditional simulators. Veins\cite{Sommer2011} is a simulation framework that connects the SUMO traffic simulator\cite{Kraj2012} to the OMNeT++ event-based communication network simulator\cite{Varga2001}. Despite its wide range of applications for the simulation of vehicular ad hoc networks, this coupled platform cannot support the investigation of many novel applications, such as Cooperative Collision Warning (CCW) systems, without requiring significant changes to the underlying software structure.

Another category of vehicular simulators has a particular focus on the simulation and analysis of different algorithms on a single vehicle. However, the main disadvantage of such simulators is the lack of complexity of an urban driving setup, which sets significant limitations on the examination of an ADAS in a large-scale traffic scenario. Such traffic-related shortcomings may include the absence of intersections, traffic rules, and most importantly, other vehicles. TORCS\cite{TORCS} is an instance of a such simulator.

With the astonishing rise and success of machine learning approaches, a new type of simulator is emerging that focuses on machine learning, specifically deep learning applications within ITS. They are, however, very similar to the preceding category in that they lack an extensive traffic setup and are primarily concerned with autonomous driving and the performance of a single, fully automated vehicle. CARLA\cite{Doso2017}, developed by researchers at Intel, and AirSim\cite{Shah2017}, developed by Microsoft, are examples of such simulators that are both open-source and are powered by Unreal Engine 4 (UE4)\cite{UE4}.

This paper's contributions to overcoming the aforementioned issues are as follows:
\begin{itemize}
    \item This paper and its companion \cite{Jami2017} propose a comprehensive co-simulation framework designed specifically for the analysis of automated applications such as vehicular safety systems within a hybrid transportation network that includes both human-driven and automated vehicles. It is made up of perfectly designed modules such as mobility and human driver models, a road transportation package, inter-vehicular communication protocols, and safety systems such as FCW.
    \item Introducing a separate driver model that allows for the simulation of dangerous road traffic scenarios while overcoming the inherent perfection and accident-free nature of a Car-Following Model (CFM).
    \item The proposed co-simulator allows for real-time representation of various traffic environments.
\end{itemize}

The technical aspects of these components are discussed in the following sections. The rest of the paper is organized as follows: 
In section \ref{sec:co-simulation_architecture}
we describe the co-simulation architecture in detail and propose a driver-in-the-loop approach to simulate human-driven vehicles. Next, we study and introduce different driving characteristics in section \ref{sec:human_driving_characteristics}. In section \ref{sec:dataset_analysis} we analyze a congested traffic dataset and perform a classification of three classes of drivers. Moreover, we extract descriptive information for different categories of drivers where we use all this information and recreate a congested traffic scenario in our simulation in section \ref{sec:results}. We conclude by analyzing various drivers' responses to FCW and their impact on the performance and safety of the traffic system. 

\section{Co-Simulation Architecture} \label{sec:co-simulation_architecture}
\noindent As the current conventional transportation system evolves rapidly into an intelligent one, the quick and efficient development of many novel ITS technologies necessitates simulation-based testing and evaluation. The co-simulation framework presented in this paper is an extension of our previous work\cite{Jami2017}, and it was created with the cutting-edge game engine technology UE4\cite{UE4} and its integrated NVIDIA PhysX Vehicle\cite{PhysX}. 
UE4 is a video game development tool created by Epic Games, a video game and software development company. Developers can use this tool to create simulations, edit videos or sounds, and render animations. Nvidia's PhysX is an open-source real-time physics engine middleware SDK that is part of the Nvidia GameWorks software suite. PhysX's vehicle support has been expanded from suspension/wheel/tire modeling to a more comprehensive model that couples modular vehicle components such as the engine, clutch, gears, auto box, differential, wheels, tires, suspensions, and chassis.

\begin{figure*}
    \centering
    \includegraphics[width=0.7\linewidth]{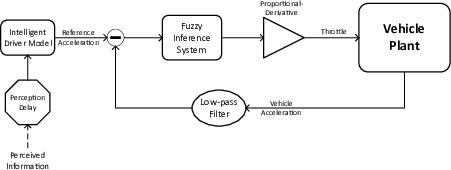}
    \caption{The Fuzzy-PD control structure used to represent a human-driver's car following behavior}
    \label{fig:controller}
\end{figure*}

We used a modular pipeline to decompose the driving task into different interoperable components, as opposed to many traditional simulators that use a CFM to handle both vehicle dynamics and driver characteristics\cite{Kraj2012, Vissim, Treiber2010}. These include a mobility model, a driver model, perception and sensing packages, and a CCW system that perceives the surrounding environment using information obtained from the vehicle's local sensors and inter-vehicular communication. This active safety mechanism in our simulation framework assists the driver in avoiding dangerous scenarios by issuing imminent alerts. Figure \ref{fig:architecture_overview} represents an overview of the proposed simulation platform and illustrates the relationship between its components. The scenario that is shown on the right side of the figure is a snapshot of the simulation environment and demonstrates vehicles equipped with CCW in a near-crash scenario
\footnote{A demo of the simulator demonstrating the evaluation and effectiveness of a CCW in avoiding a crash can be found at \url{https://vimeo.com/252441087}}.

\subsection{Mobility Model}
\noindent The mobility model is responsible for handling the kinematics and dynamics of a vehicle and its subsystems. This module converts the driver-supplied throttle and steering angle values into lateral and longitudinal vehicle movement. We use the UE4 standard physics engine and its integrated component, NVIDIA PhysX Vehicle, to handle the physical characteristics of vehicles in 3D environments. The NVIDIA PhysX models a vehicle as a rigid body on a collection of sprung masses\cite{PhysX}. The complete UE4 physical engine allows for a near-realistic dynamical response of vehicles within its virtual environment by adjusting the vehicle's mechanical setup, such as engine torque and differential setup, and by incorporating additional components such as tire models and road friction.
\subsection{Driver Model}
\noindent The driver model, which represents a human driver and is tasked with controlling the vehicle while adhering to traffic laws, is essentially the system's brain. A driver controls the vehicle during a longitudinal car-following regime by adjusting the throttle and brake pedals to maintain a safe speed and distance from the leading vehicle\cite{9837992}. According to control theory, the interaction between the driver and the vehicle creates a closed-loop feedback control system, with the driver acting as a feedback controller\cite{Filev2011}. As the most popular type of controller in the industry, PID controllers are often used to describe robot driver models\cite{Filev2011, Burnham1974}. Despite their popularity and ease of implementation, PID-like models alone cannot characterize ideal human drivers because they are based solely on speed and acceleration errors and do not take into account driver-specific characteristics. A more accurate representation of a driver should consider nonlinearities in human response time, as well as their specific decision processes\cite{Zheng2005}.

Given that each driver has a unique driving style, we propose a mixed-driver model that is controlled by a CFM-Fuzzy-PD closed-loop feedback controller. As an alternative to PD control, fuzzy-PD control is often mentioned because some of the techniques and algorithms used to tune the PD gains require further tuning by a skilled human driver. The combination of these models allows us to incorporate the most beneficial features of each subsystem, resulting in a natural and human-interpretable system, suitable for implementing a driver-in-the-loop model. The CFM interprets the data and recommends a target acceleration. The Fuzzy-PD components then determine how quickly to adapt to the provided acceleration value.

A Fuzzy Inference System (FIS) is a viable option for dealing with the uncertainties in the car-following regime by mapping the variables that the driver perceives to the variables that they can control with arbitrary accuracy using fuzzy reasoning. The Partial-Derivative (PD) component stabilizes the response time and ensures the controller's performance. The proposed CFM-Fuzzy-PD closed-loop feedback controller is depicted in Figure \ref{fig:controller}.

We begin our discussion of our driver-in-the-loop system with the Intelligent Driver Model (IDM) component. In our system, IDM is a microscopic model of traffic flow that provides the reference acceleration $a_i(t)$ to the driver's fuzzy component\cite{Treiber2000}. The microscopic traffic model's goal is to describe the dynamics of each vehicle as a function of the positions and velocities of its neighbors\cite{Panwai2005}.

The output of such a model, which can be either the vehicle's velocity or acceleration, can be interpreted as the desired velocity/acceleration of the driver while following other vehicles in a smooth and safe driving regime. The IDM was designed by \cite{Treiber2000}, which is also implemented in several microscopic traffic simulators such as MovSim \cite{Treiber2010} and SUMO \cite{Kraj2012}. The model is formulated as follows:
\begin{equation}
    v_i(t) = \frac{dx_i(t)}{dt}
\end{equation}
\begin{equation}
    a_i(t) = \frac{dv_i(t)}{dt} = \alpha(1 - (\frac{v_i(t)}{v_0})^\delta - (\frac{G(t)}{s_i(t)})^2)
    \label{eq:idm}
\end{equation}
with 
\begin{equation}
    G(t) = s_0 + v_i(t)\tau + \frac{v_i(t)\Delta v_i(t)}{2\sqrt{\alpha \beta}}
\end{equation}
where $x_i(t)$, $v_i(t)$, and $a_i(t)$ represent the position, velocity, and acceleration of the host vehicle at time $t$ respectively; $\alpha>0$ is the comfortable acceleration of the driver, $\beta>0$ represents comfortable braking deceleration; $v_0$ is the desired velocity; $s_0$ represents the minimum gap; $s_i(t)=x_{i-1}-x_i-l_{i-1}$ is the net distance between host vehicle $i$ and leader vehicle $i-1$ with a length of $l_{i-1}$; $\tau$ is the minimum desired safe time headway to the vehicle in front; and $\Delta v_i(t)=v_i-v_{i-1}$ represents the velocity between the host and leader vehicles.

Although IDM includes some of the most important driver characteristics, it only provides the reference acceleration
$a_i(t)$ as a desired acceleration value for the driver. It is up to the rest of the controller to figure out how fast the driver wants to reach that value. Given a complete dynamical model of a vehicle, the Fuzzy-PD part of the controller attempts to map the supplied acceleration values to changes in the throttle and brake pedals. The statistical analysis in the work of \cite{Xu2015} shows that the integral of throttle position over time has a direct relationship with fuel consumption, and thus needs to be considered in modeling and identification of human drivers. However, due to the lack of reliable and publicly available traffic data that include throttle measurement, we constructed the Mamdani FIS component using a heuristic and human interpretable approach. As a result, the FIS in this case is designed to be as simple yet interpretable as possible, with only one input $\Delta a_i(t)$ denoting the difference in accelerations and one output $\delta p$ denoting changes in accelerator or brake pedal positions. The seven linguistic variables of this Mamdani FIS are shown in Figures \ref{fig:input_membership_functions} and \ref{fig:output_membership_functions}.
The fuzzy rules are characterized in a way that the normalized Membership Functions (MF) are within the interval $[-1,1]$. These values are then mapped to the maximum acceleration and deceleration values of the vehicle itself. Figure \ref{fig:output_fuzzy_system} shows the output of the FIS for the MFs' characteristics represented in Figures \ref{fig:fuzzy_system}a-b. These MFs are tuned intuitively to represent different driving behaviors. It is found that the steeper the slope of the curve in Figure \ref{fig:output_fuzzy_system}, the faster and more aggressively a driver would respond to acceleration changes dictated by IDM.

Many car-following models including IDM do not consider the driver perception-reaction time (time taken by a driver to sense the stimulus and start reacting by either accelerating or decelerating) \cite{Panwai2005}. Perception-reaction time is a driver-specific parameter that can be influenced by a variety of implicit and explicit factors such as the driver's mood, drowsiness, comfort level, vehicle condition, and weather. Olson \etal \cite{Olson1986} conducted a study on human reaction time and found that the average perception-reaction time for $95^{th}$ percentile was $1.6$ seconds. Similarly, we adopt a constant perception-reaction time of $1.4s$ and a constant pedal switch time of $0.2s$.

\begin{figure}
    \centering
    \includegraphics[width=0.8\linewidth]{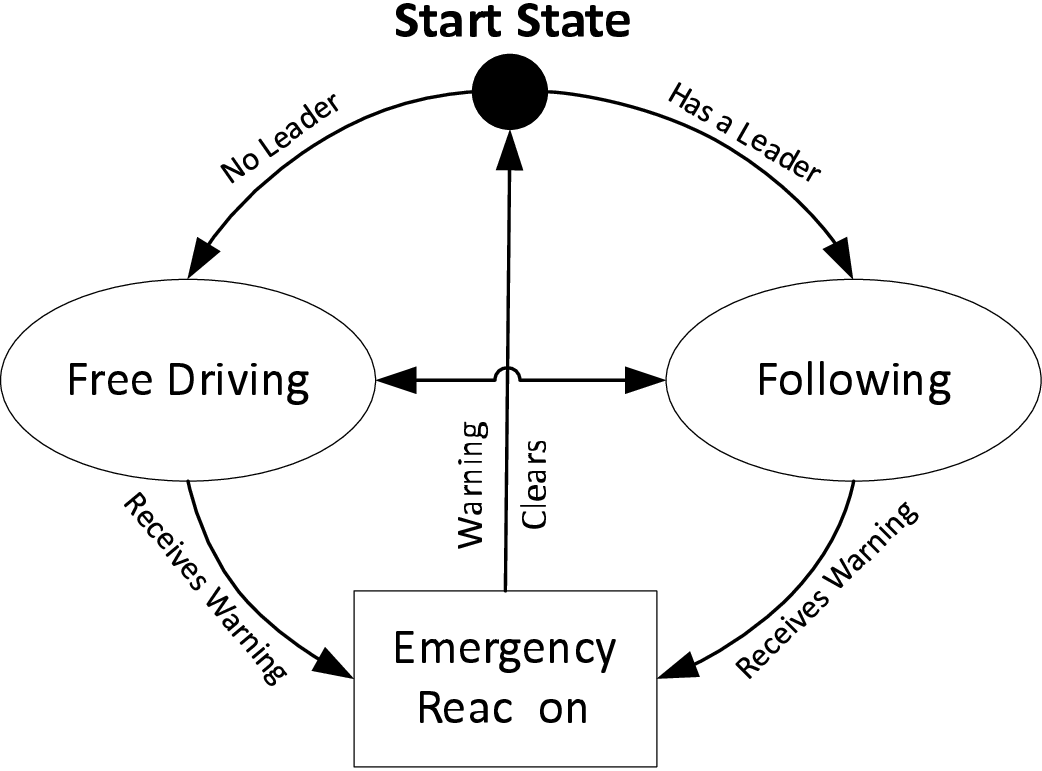}
    \caption{Main driving tasks represented through a state machine}
    \label{fig:driving_tasks}
\end{figure}

A low-pass filter is the final component of this feedback control system. Because the dynamical behavior of the vehicle is represented by a complete 3D physics model, factors other than throttle value affect the vehicle's acceleration. Tire and suspension models, as well as road characteristics, are examples of these. The moving average filter used here helps to stabilize the system's response by reducing noise artifacts of the vehicle's acceleration value. Finally, we disintegrated the IDM into individual states and introduced a driver state parameter that identifies the appropriate CFM state for the driver. The driver's state is determined by information gathered from various sources, such as vision and safety alerts. A field-of-view represents the driver's vision, allowing it to sense the objects within it and schedule an appropriate action accordingly. These actions actually represent the states that the driver can enter. Aside from vision, other factors such as safety warnings may cause a state transition and thus force the driver into another state. Figure \ref{fig:driving_tasks} depicts the main driving tasks in the form of a state machine.

\begin{figure}
     \centering
     \begin{subfigure}[b]{0.45\textwidth}
         \centering
         \includegraphics[width=\textwidth]{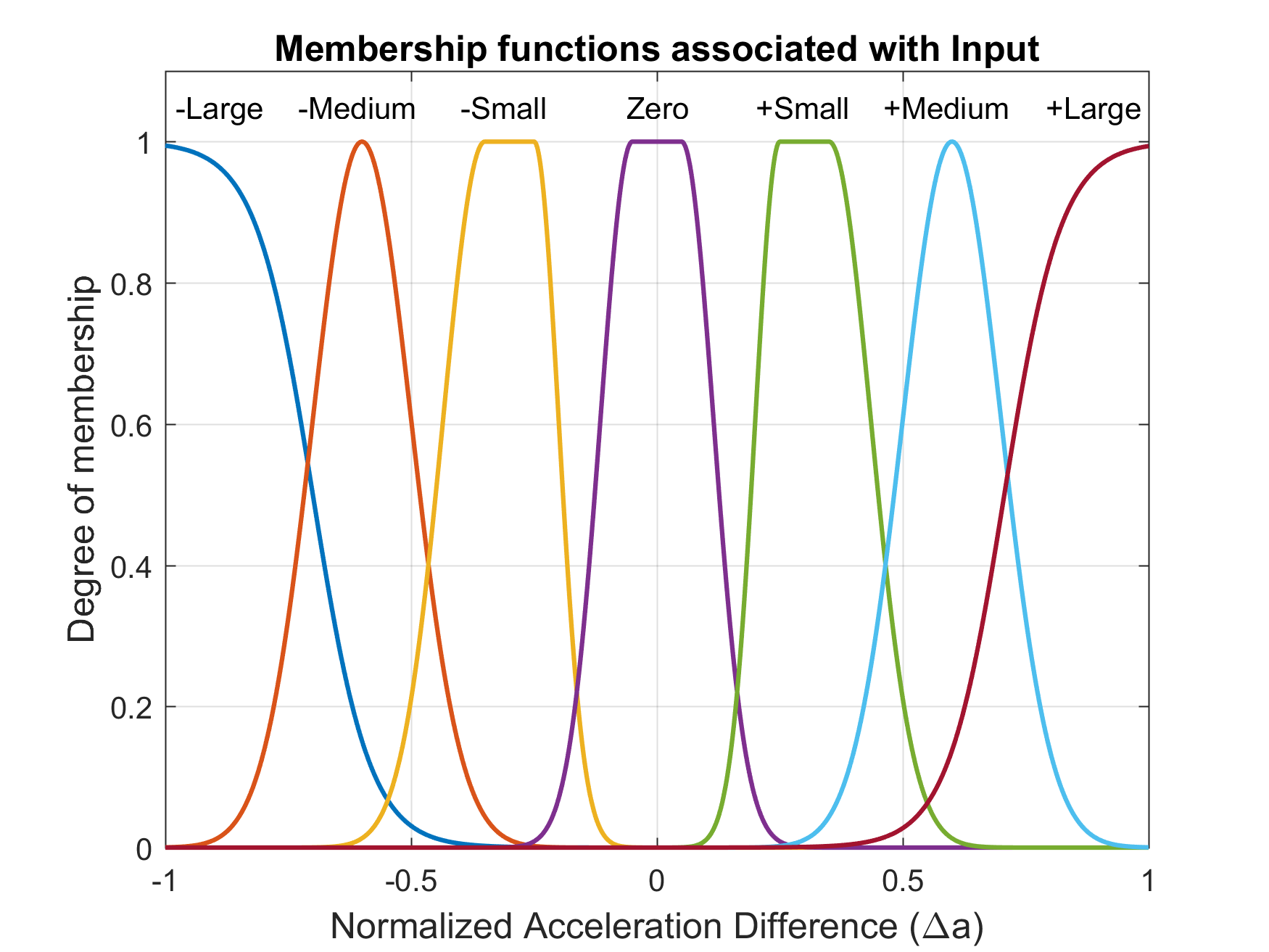}
         \caption{Membership functions associated with inputs}
         \label{fig:input_membership_functions}
         \vspace{-0.2in}
     \end{subfigure}
\end{figure}
\begin{figure}\ContinuedFloat
     \begin{subfigure}[b]{0.45\textwidth}
         \centering
         \includegraphics[width=\textwidth]{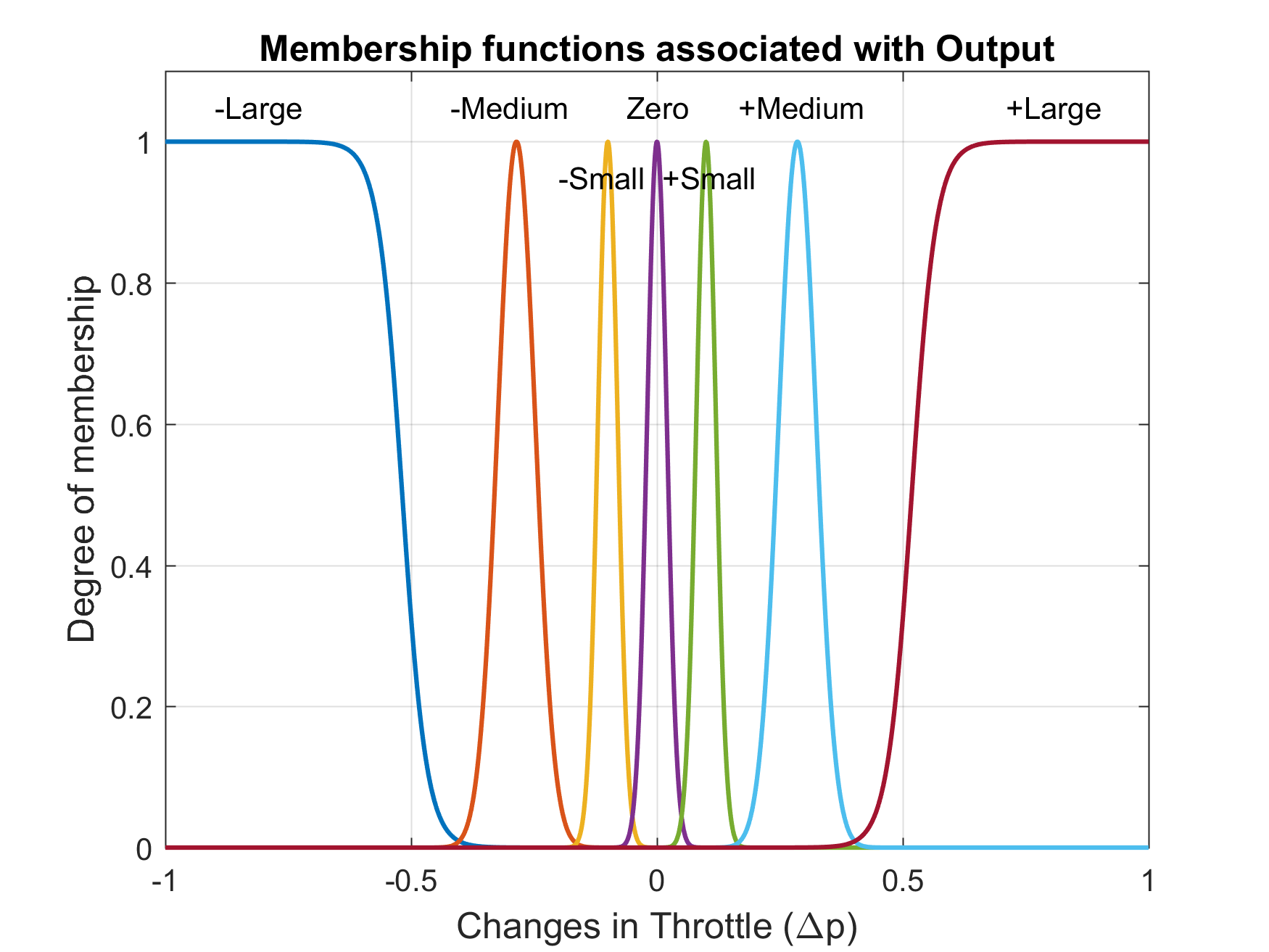}
         \caption{Membership functions associated with outputs}
         \label{fig:output_membership_functions}
         \vspace{-0.2in}
     \end{subfigure}
\end{figure}
\begin{figure}\ContinuedFloat
     \begin{subfigure}[b]{0.45\textwidth}
         \centering
         \includegraphics[width=\textwidth]{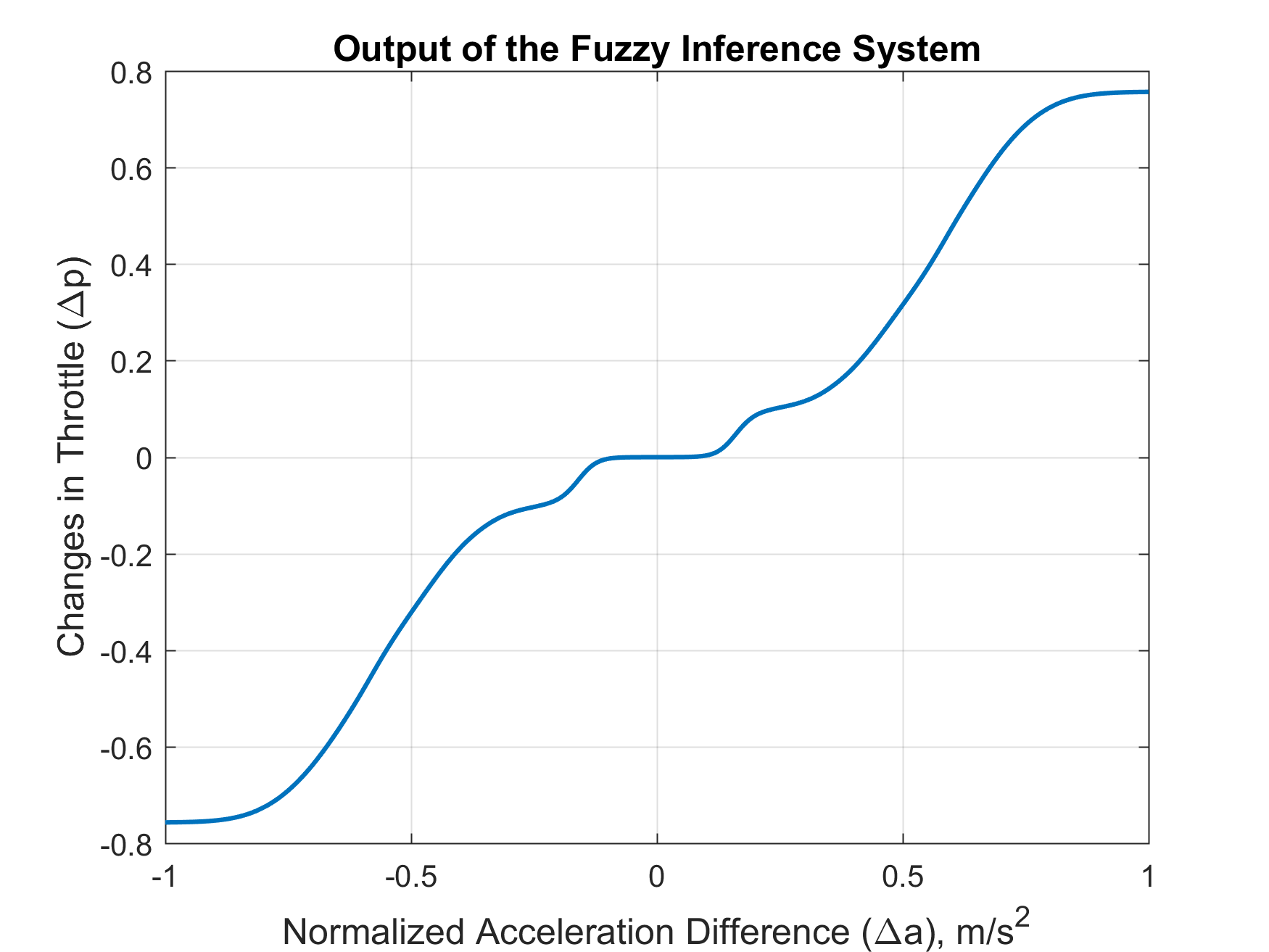}
         \caption{Output of the fuzzy inference system for above membership functions}
         \label{fig:output_fuzzy_system}
     \end{subfigure}
        \caption{Fuzzy inference system membership functions}
        \label{fig:fuzzy_system}
        \vspace{-0.2in}
\end{figure}

A visual representation of changes in vehicle kinematic data is shown below using the proposed driver-in-the-loop model for each state of the IDM (Figure \ref{fig:kinematic_response}). In Figures \ref{fig:kinematic_response_free_flow} and \ref{fig:kinematic_response_following_state} the free flow and following states are also compared with the case of traditional simulators approach, i.e., a pure CFM. Figure \ref{fig:kinematic_response_emergency_state} shows the host vehicle's reaction in a near-crash scenario where the leading vehicle exhibits sudden braking behavior. As a final note, the parameters of CFM-Fuzzy-PD system are tuned to represent a moderate driving behavior.

Finally, the introduction of a separate driver model in such a modular manner lays the foundation for a simple, human-interpretable system that can be tuned to represent various driving characteristics. Additionally, it breaks down the traditional one-to-one approach of CFMs and allows for the introduction of different classes of drivers.

\subsection{Inter-Vehicular Communication Model}
\noindent The safety algorithms in CVS Systems use the communicated data between vehicles to survey the thread zones every $0.1s$ to detect any danger. Cellular-V2X, also known as C-V2X, is the leading standard candidate for inter-vehicular communication among the suggested communication protocols \cite{8628416}.
C-V2X has been specifically designed for ITS applications as a fast-response and scalable solution due to the time sensitivity of active safety systems and the need for low latency in their communication protocols. It is also worth noting that the mandatory deployment of C-V2X devices on all vehicles is currently being debated at the United States Department of Transportation.

\textbf{Communication Logic:} This component represents the communication subsystem on the host vehicle. It determines the transmission power and frequency and dictates how the recourses of the network should be used to broadcast the next Basic Safety Message (BSM)\footnote{A basic safety message holds essential vehicle's state information such as its location and kinematic data \cite{J2735}.}. The choice of communication parameters directly affects the reception probability of BSM packages within a V2V environment \cite{Fallah2011, J2735}. The baseline transmission rate in CVS Systems  is $10Hz$ \cite{DOT_HS_810_591}.

\textbf{Vehicle Tracking:} This module deals directly with the tracking of vehicles whose information is received over the communication medium. Upon reception of BSMs by the host vehicle, a local map is constructed from these communicated data which will keep track of the surrounding vehicles. Various methods have been proposed to minimize the tracking error. A few notable ones include simple models such as first and second order kinematic models (constant speed and constant acceleration), to more complex ones that are based on Kalman Filter and Gaussian Process\cite{9644629, razzaghpour2023controlaware, 10076814}.
Nevertheless, it has been shown that in a longitudinal car-following regime, a second order kinematic model can perform with high accuracy \cite{Khandani2017}. Consequently, the constant acceleration is also the model we adopt in our framework.

\textbf{Effect of Communication Layers:} In a communication network model, the application layer perceives the effect of the lower layers on the communicated data as a pattern of random losses imposed on the stream of packets that are received by the host vehicle. Knowing that the transmission delay in a single hop wireless network is insignificant and that the losses in a broadcast network are well-randomized, we can describe the effect of the communication network in an abstract mathematical form using a random loss rate during transmission \cite{Fallah2011, Fallah2015, 9773892}.

\begin{figure}
     \centering
     \begin{subfigure}{0.42\textwidth}
         \centering
         \includegraphics[width=1.02\textwidth,trim={5mm 10mm 10mm 5mm},clip]{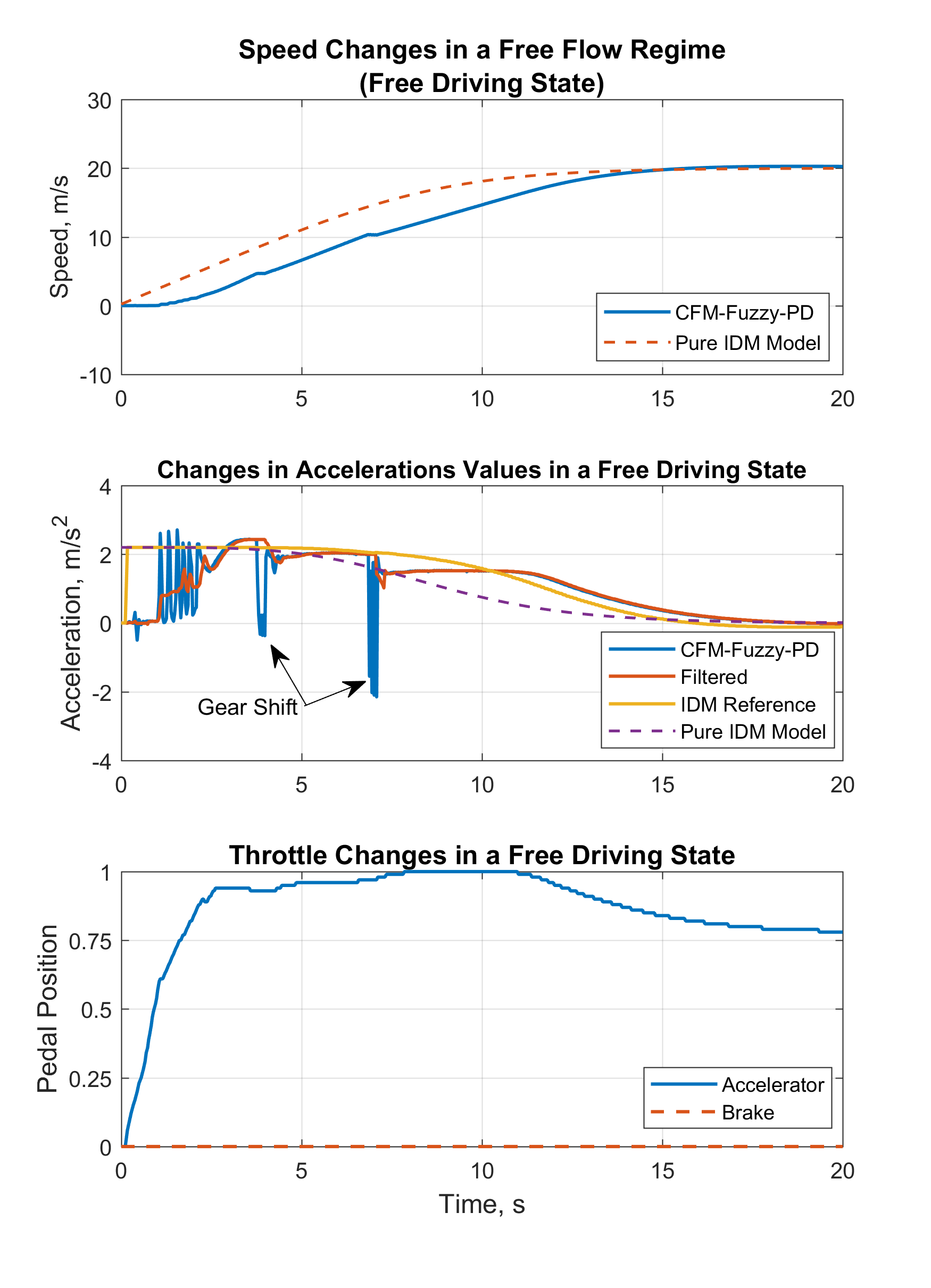}
         \caption{Free Flow}
         \label{fig:kinematic_response_free_flow}
         \vspace{-0.2in}
     \end{subfigure}
\end{figure}
\begin{figure}\ContinuedFloat
     \begin{subfigure}{0.42\textwidth}
         \centering
         \includegraphics[width=1\textwidth,trim={5mm 10mm 10mm 5mm},clip]{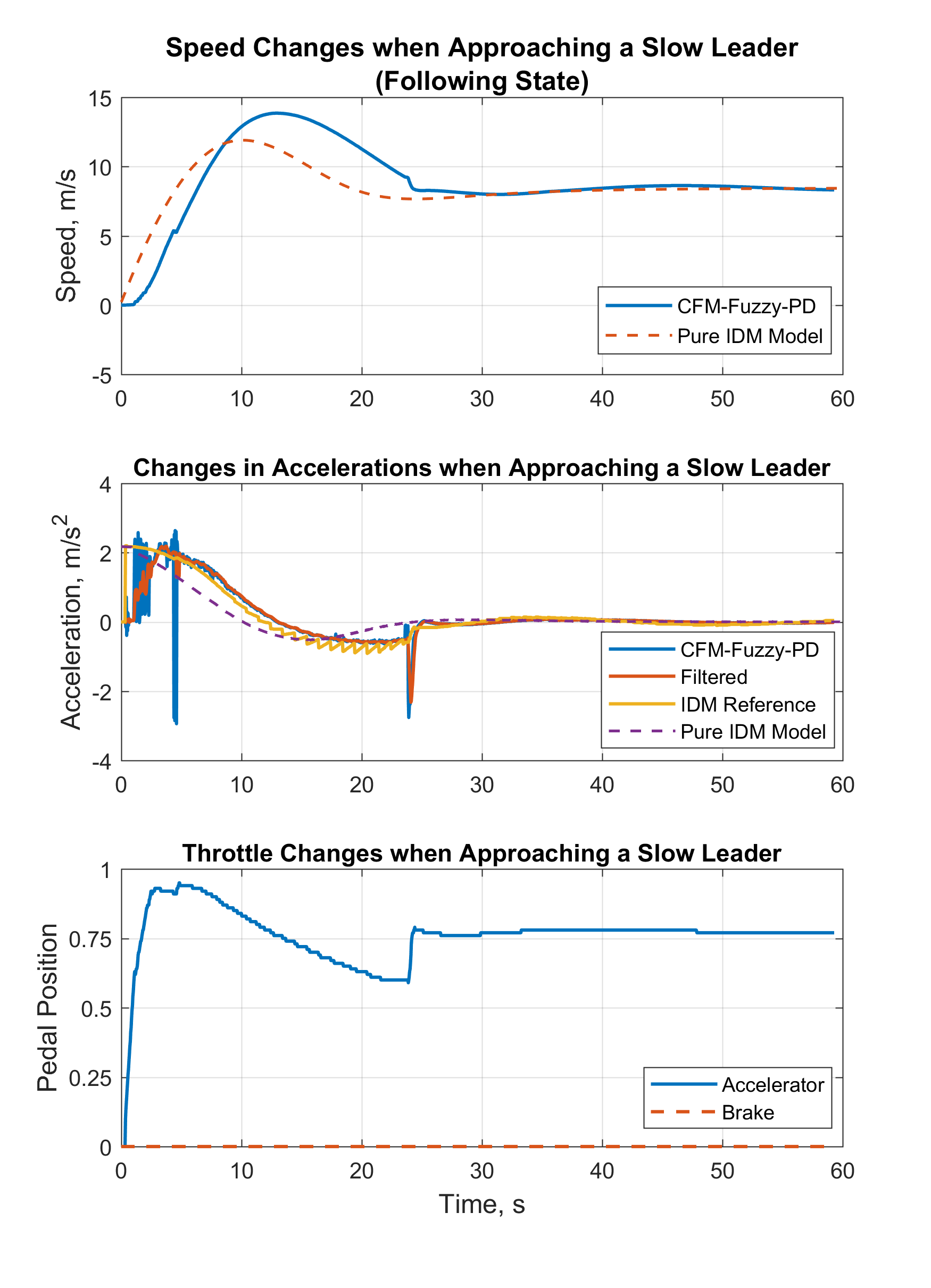}
         \caption{Following}
         \label{fig:kinematic_response_following_state}
         \vspace{-0.2in}
     \end{subfigure}
\end{figure}
\begin{figure}\ContinuedFloat
     \begin{subfigure}{0.41\textwidth}
         \centering
         \includegraphics[width=1.02\textwidth,trim={5mm 10mm 10mm 5mm},clip]{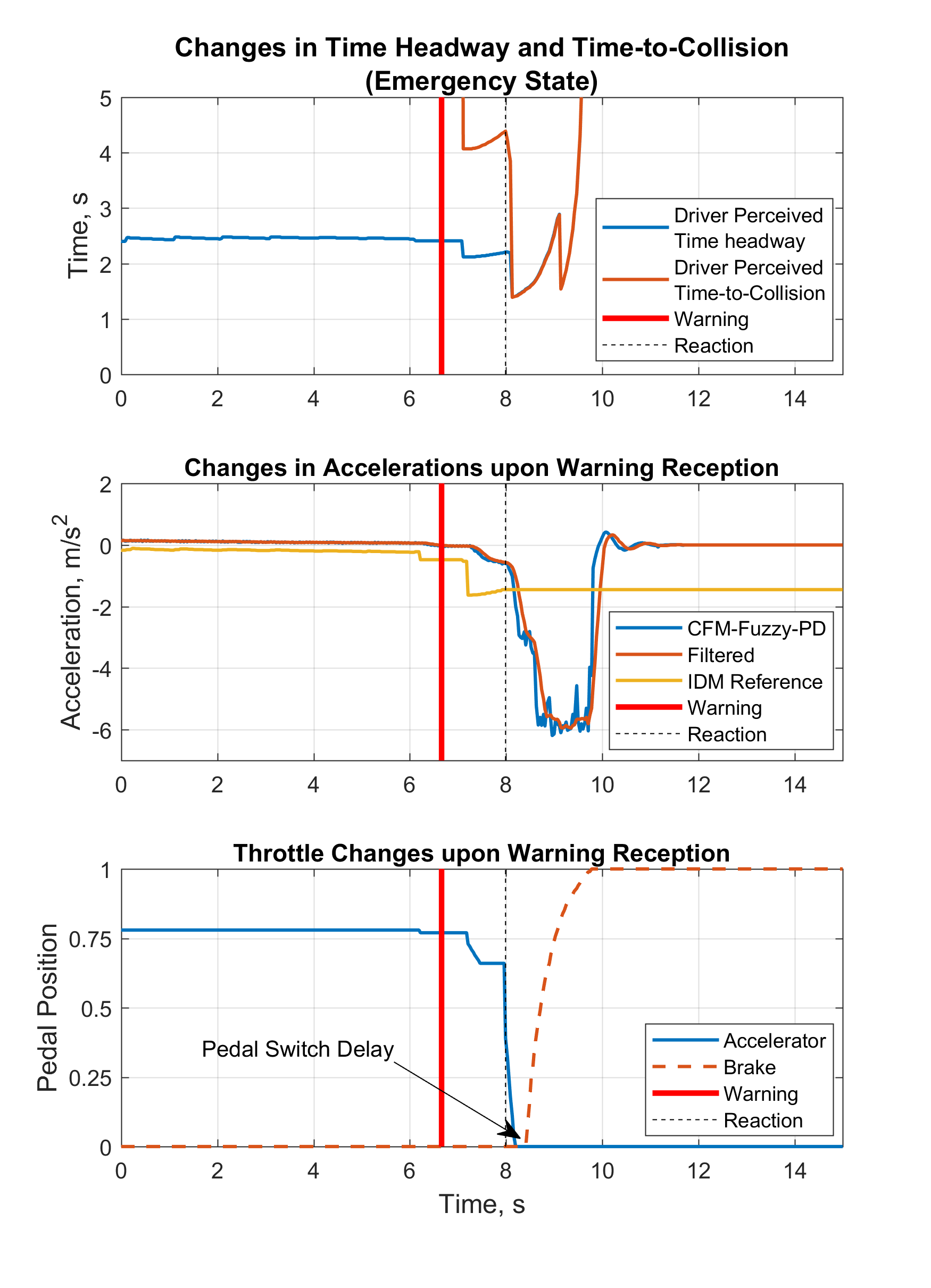}
         \caption{Emergency state}
         \label{fig:kinematic_response_emergency_state}
     \end{subfigure}
    \caption{A visualization of the kinematic responses of the proposed driver-in-the-loop model for three main driving states (a) Free Flow, (b) Following, and (c) Emergency state.}
    \label{fig:kinematic_response}
    \vspace{-0.1in}
\end{figure}

\subsection{Forward Collision Warning Algorithms}
\noindent One of the most important collision prevention mechanisms is FCW. CVS systems can be a low-cost way to integrate FCW systems into existing vehicles. The exchange of BSMs provides the host vehicle with rich content for complex predictions. Various automakers are testing and using various FCW algorithms. Unfortunately, most automakers do not disclose the details of their safety algorithms, so we can only rely on those available to academic and research communities. 
In our companion paper\cite{Jami2017}, we analyzed the performance of some of such FCW algorithms, namely the Knipling algorithm\cite{Knipling1993}, CAMP Logistic Regression\cite{Kiefer2003}, and a driver-tuned FCW algorithm from National Highway Traffic Safety Administration (NHTSA) \cite{Brunson2002}. In this work, we focus on only two of these algorithms that showed better performance, i.e., CAMP Logistic Regression and NHTSA.

\section{Understanding Human Driving Characteristics} \label{sec:human_driving_characteristics}
\noindent Human factors play an important role in any cyber-physical system that includes humans as part of its physical plant, such as drivers and pedestrians in an ITS. The past decade has seen a general downward trend in road fatalities, thanks to the advancement in safety systems of modern vehicles. However, crash statistics are still alarming, as the recent NHTSA study shows a total of 42,915 lives lost in 2021 alone. NHTSA study shows a total of 37,133 lives lost in 2017 alone, of which 3,166 cases involved distracted driving \cite{NCSA2017}. Another summary of driving statistics shows that up to 660,000 drivers use cell phones while driving during daylight hours, increasing their chances of being involved in an accident\cite{NCSA2013}. The study states that approximately 80\% of recorded crashes were the result of drivers' distraction, making it the primary cause of rear-end crashes. However, the driver's inattention does not necessarily imply the carelessness of that driver\cite{Knipling1993}. A driver's attention may be diverted from the forward path for both driving and non-driving reasons. Examples of driving-related distractions may include watching for a pedestrian, looking at signs and landmarks, looking at side mirrors, or turning the head during a lane change. 

Despite the importance and the need for detailed study, distracted driving does not necessarily describe a specific driver's driving personality. Instead, it defines a state in which the driver is no longer paying attention to the driving task. In other words, one cannot bind distracted driving to their driving characteristics (as this does not relate to how the kinematics of vehicles are affected). On the other hand, there exists a category of dangerous on-the-road driving behavior that is labeled "aggressive driving." NHTSA defines aggressive driving as "the operation of a motor vehicle in a manner that endangers or is likely to endanger persons or property"\cite{Stuster2004}. An aggressive driver may exhibit behaviors such as speeding, failing to obey traffic signals, making frequent lane changes, and most importantly, tailgating which is considered one of the significant causes of crashes that may result in severe injury or death. Recent NHTSA reports\cite{NCSA2017} show that speeding, which is a characteristic of aggressive drivers, has been the cause of approximately 9,717 deaths, accounting for more than 26\% of all traffic fatalities in 2016 alone.

These alarming statistics compel the implementation of active vehicular safety systems, which brings us back to the goal of this paper: introducing a simulation platform for the evaluation and verification of vehicular active safety systems in various traffic scenarios.
The simulation architecture described in this paper allows for the introduction of two types of drivers, namely cautious and distracted, as well as three distinct driving behaviors, namely normal, conservative, and aggressive. As a result of this segregation, a driver may be in any of six different states at any given time. We use a driver's visual perception to determine their cautious and distracted states. A cautious driver is one who pays close attention to the road ahead and what is in front of him/her, whereas a distracted driver is one whose vision range is severely limited, causing him/her to be unaware of his/her surroundings. A distracted driver has a very high chance of colliding with a decelerating Leading Vehicle (LV) in a congested traffic scenario or at a traffic light in such a system.

The driving characteristics of a driver can be extracted from a congested traffic scenario, which is considered a major contributing factor to aggressive driving\cite{NCSA2013}. Impatient drivers who are frustrated by on-the-road delays caused by either high traffic volume during rush hour or a collision may respond aggressively by either following too closely or changing lanes frequently. For this purpose, we use the Next Generation Simulation (NGSIM) dataset\cite{NGSIM} to find appropriate parameters for the IDM car-following model that will represent the aforementioned categories of drivers. The NGSIM dataset contains the vehicle paths of over 2000 drivers during an afternoon rush hour. As a result, the dataset is an excellent candidate for the task at hand. The details of the NGSIM dataset, as well as its statistical analysis, are discussed in the following section.
\section{Dataset Analysis: Parameter Extraction and Driver Classification} \label{sec:dataset_analysis}
\noindent The Next Generation SIMulation (NGSIM) program was originated by the Federal Highway Administration of the U.S. Department of Transportation (USDOT) to "improve the quality and performance of simulation tools, promote the use of simulation for research and applications, and achieve wider acceptance of validated simulation results"\cite{NGSIM}. The NGSIM dataset provides high resolution ($10 Hz$) positional and kinematic data for 5648 vehicles, which were collected during afternoon rush hour in three 15-minute intervals ($4:00$ pm to $4:15$ pm, $5:00$ pm to $5:15$ pm, and $5:15$ pm to $5:30$ pm). These vehicle trajectory data were collected on a sub-second basis on US Interstate 80 in California on April 2005, using seven stationary cameras mounted on a 30-story building. The dataset is regarded to contain the most useful features for the development and validation of microscopic traffic models. The NGSIM dataset was published in 2007 and is freely available for download. 

\begin{figure*}
     \centering
     \begin{subfigure}{0.45\textwidth}
         \centering
         \includegraphics[width=1.1\textwidth]{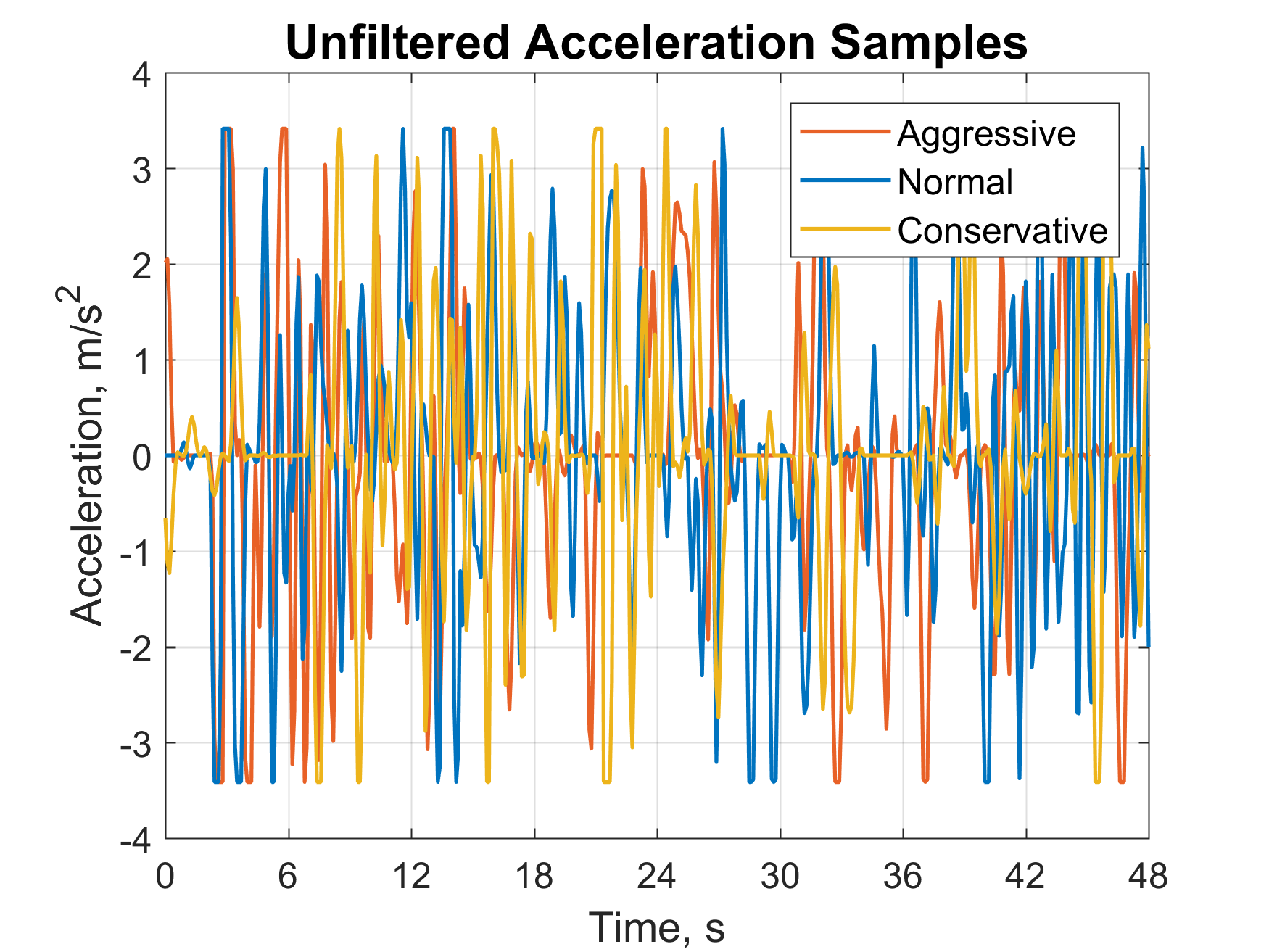}
         \caption{Unfiltered acceleration}
     \end{subfigure}
    \hfill
     \begin{subfigure}{0.45\textwidth}
         \centering
         \includegraphics[width=1.1\textwidth]{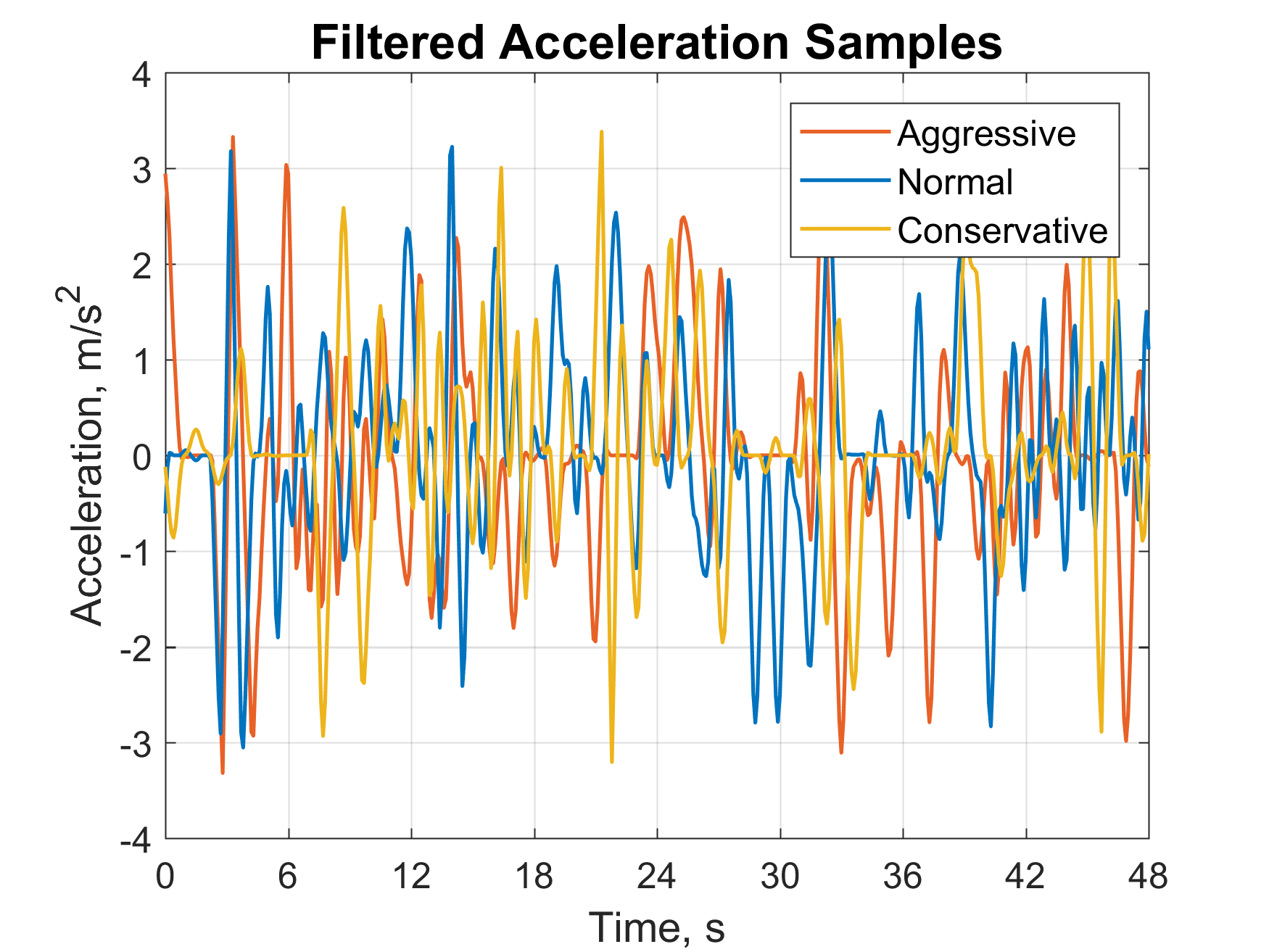}
         \caption{Filtered acceleration}
     \end{subfigure}
    \caption{Comparison of filtered and unfiltered acceleration data}
    \label{fig:acceleration_data}
\end{figure*}

Before starting our analysis, it is worth mentioning that the NGSIM dataset is provided in imperial units, which were converted to the metric system for the results reported here. Additionally, the trajectory and kinematic data of vehicles exhibited noise artifacts that were filtered using a moving average filter with a half-second window. The comparison of filtered and unfiltered sample acceleration data for three different drivers is shown in Figure \ref{fig:acceleration_data}.

Given the specification of an aggressive driver, i.e., speeding and tailgating, two main kinematic features that can best describe these aggressive characteristics are acceleration and the time headway of vehicles. Time headway is a measurement of the time it would take the following vehicle to reach the leading vehicle if the leader remains stationary.
\begin{equation}
    \tau = \frac{s_i}{v_i} = \frac{x_{i-1} - x_i - l_{i-1}}{v_i}, \ \forall v_i > 0
\end{equation}
It has been shown that drivers tend to keep a constant time headway from their leading vehicle during a car-following driving regime\cite{Helbing2001}. The probability distribution function of this feature for all drivers within the dataset is represented in Figure \ref{fig:ngsim_graphs_pdf_time_headways}. NHTSA gives several rules of thumbs and recommendations for a safe time headway. For a scenario such as the one in the NGSIM dataset (dry and clean pavement), and for the average speed of $35 mph$ ($~55 km/h$), the recommended following distance rule is $2-3$ seconds. This can also be observed in Figure \ref{fig:ngsim_graphs_pdf_time_headways} with the mode of the histogram sitting around $2.5s$. Obviously, many drivers follow this rule in an attempt to keep a safe driving experience. However, we can see that a fraction of drivers have kept a much smaller time headway to their leading vehicles, while the other group of drivers has chosen to keep larger time headway. Using the recommended time headway rule, we classify these drivers into 3 clusters: aggressive drivers with $(\tau < 2.0s)$, normal drivers $(2.0s \leq \tau \leq 3.0s)$, and conservative drivers $(3.0s < \tau)$. The changes in time headway of three randomly sampled drivers from all bins, along with the mean of their time headways are represented in Figure \ref{fig:ngsim_graphs_sampled_time_headways}. We can see that these drivers attempt to keep a relatively constant time headway all the time.

After categorizing these drivers into three groups, we attempt to extract the IDM parameters using statistical analysis of time headway and acceleration data. The average time headway PDF contains enough information to determine the distribution ratio of the drivers and their minimum desired time headway $\tau$ (equation \ref{eq:gamma_probabiliy_distribution}). We use a gamma probability distribution function of the following form to fit this data:
\begin{equation}
    f_\Gamma (x | \alpha, \beta) = \frac{\beta ^ \alpha}{\Gamma (\alpha)}x^{\alpha - 1}e^{-\beta x}
    \label{eq:gamma_probabiliy_distribution}
\end{equation}
thus, representing the time headway $\tau$ of each driver as a gamma-distributed random variable $x$ with shape parameter $\alpha=9.15$ and rate parameter $\beta=0.31$. Based on the above classification, and the fitted gamma PDF, the ratio of each category of the driver can be obtained using the inverse percentile at the edge points 2 and 3 seconds. These ratios are presented in table \ref{tab:idm_parameters}. A simulated driver will be assigned a category based on bins' thresholds and the randomly gamma-distributed time headway.

Next, we analyze the changes in the acceleration profiles to find the remaining parameters of IDM. Figure \ref{fig:ngsim_graphs_ecdf_accelerations} represents the Empirical Cumulative Distribution Function (ECDF) of all the acceleration values of each driver's category. In this figure, the solid lines represent the acceleration values while the dashed lines represent the absolute values of decelerations for each class of driver. We can see that the aggressive driver has kept higher acceleration values than the other two drivers. As we move towards the conservative drivers, the resistance in accelerating/decelerating becomes more apparent.
From this observation, the comfortable accelerations $\alpha$ and braking decelerations $\beta$ (equation \ref{eq:idm})
were represented as a uniformly distributed random variable within the range of $70^{th}$ to $90^{th}$ percentile of accelerations and decelerations respectively. The final extracted ranges for each category of drivers are shown in table \ref{tab:idm_parameters}.

\begin{figure}
     \centering
     \begin{subfigure}[b]{0.45\textwidth}
         \centering
         \includegraphics[width=\textwidth]{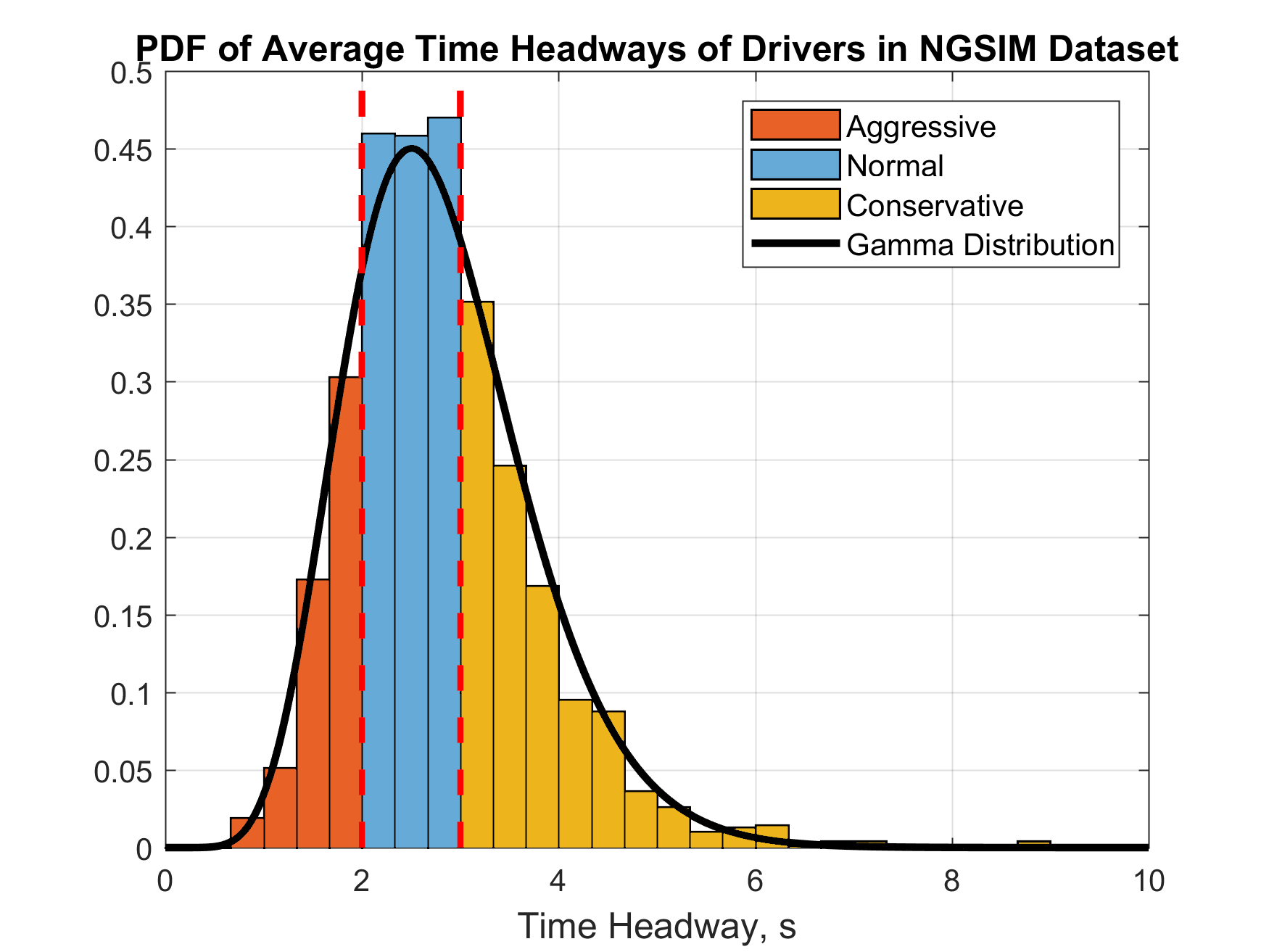}
         \caption{Distribution and clustering of mean time-headways of all drivers in NGSIM dataset}
         \label{fig:ngsim_graphs_pdf_time_headways}
     \end{subfigure}
\end{figure}
     
\begin{figure}\ContinuedFloat
     \begin{subfigure}[b]{0.45\textwidth}
         \centering
         \includegraphics[width=\textwidth]{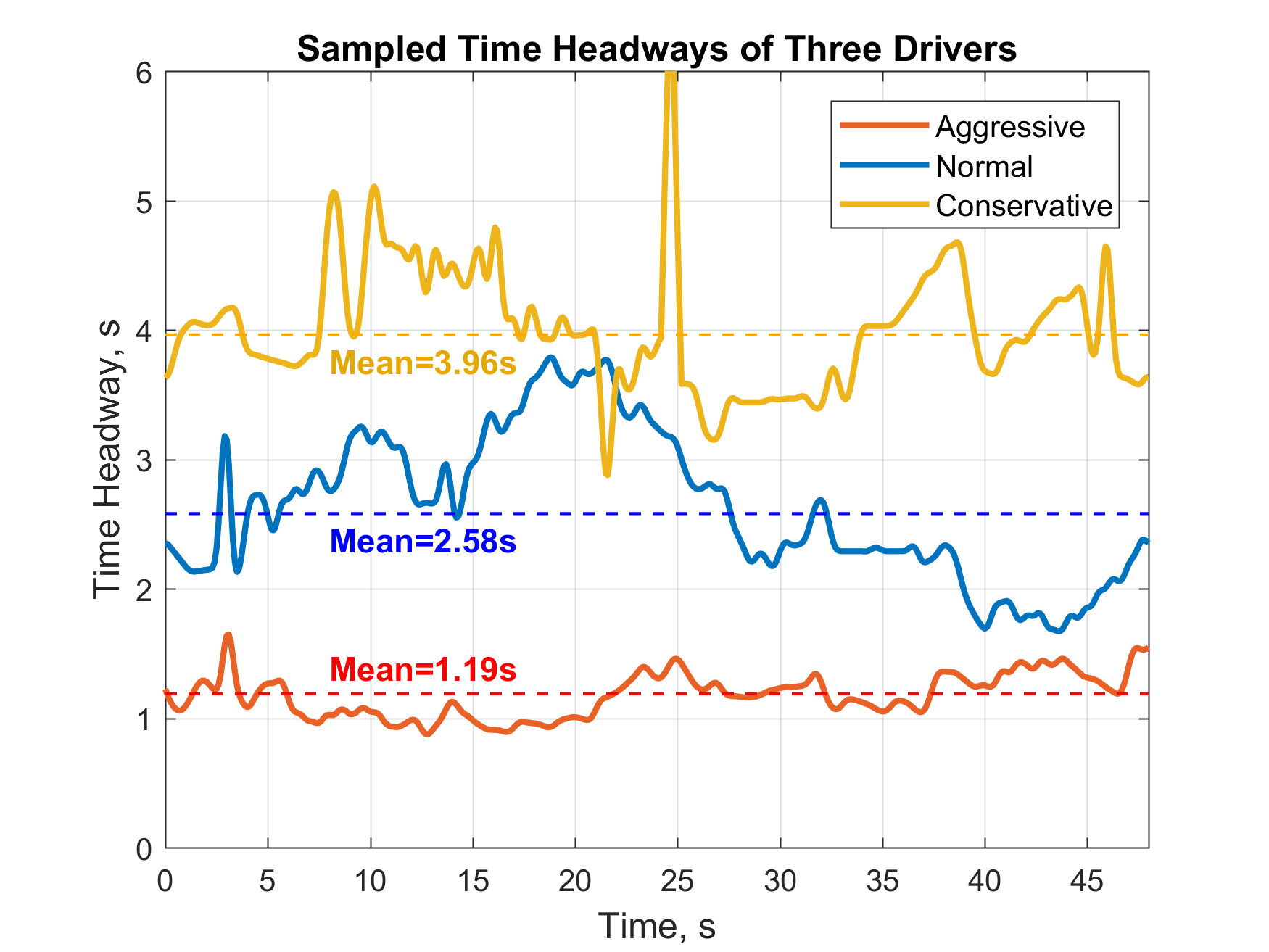}
         \caption{Changes in time headway of three sampled drivers}
         \label{fig:ngsim_graphs_sampled_time_headways}
     \end{subfigure}
\end{figure}

\begin{figure}\ContinuedFloat
     \begin{subfigure}[b]{0.45\textwidth}
         \centering
         \includegraphics[width=\textwidth]{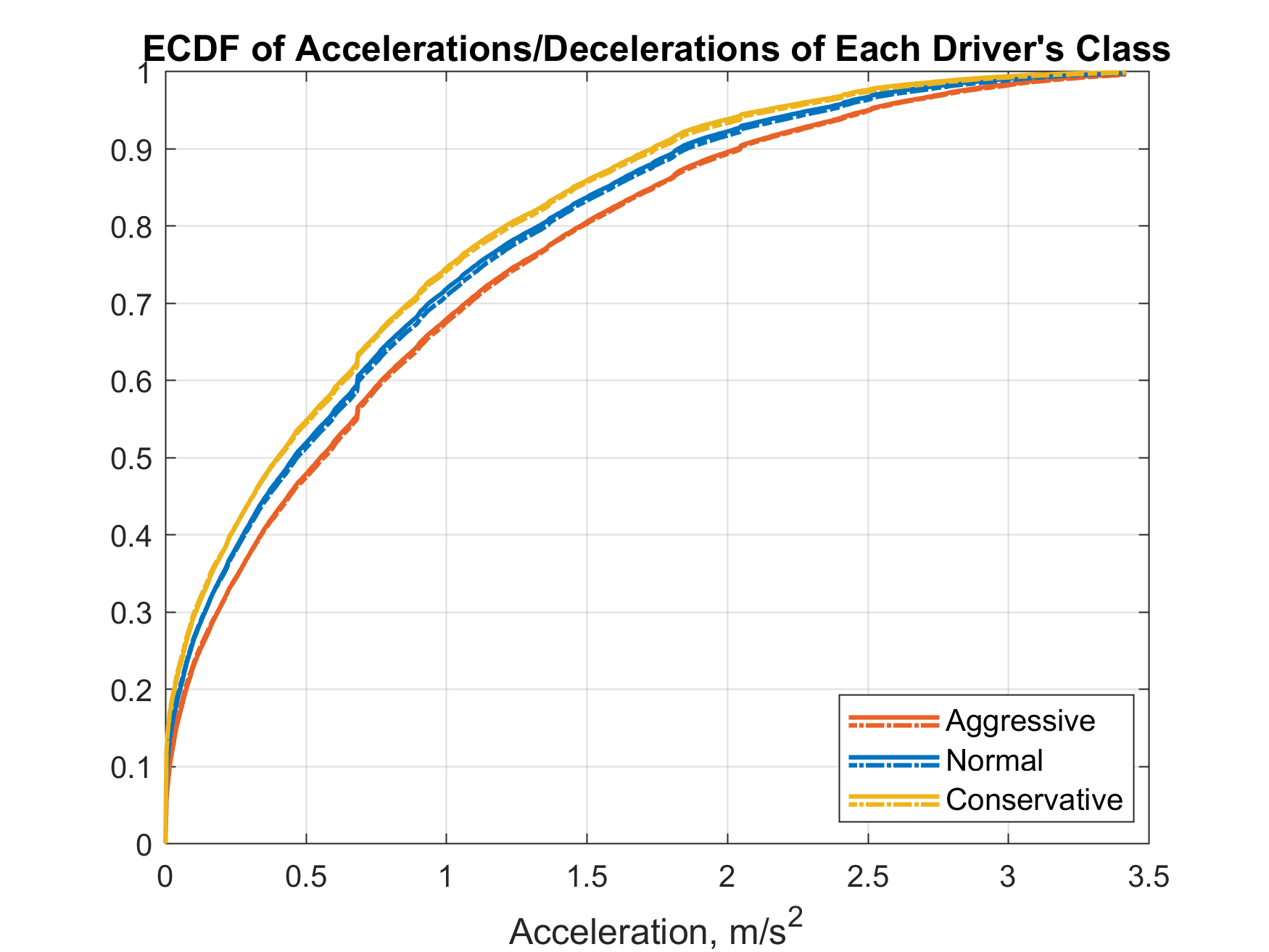}
         \caption{ECDF of acceleration values of all drivers of each category}
         \label{fig:ngsim_graphs_ecdf_accelerations}
     \end{subfigure}
        \caption{NGSIM dataset analysis}
        \label{fig:ngsim_graphs}
\end{figure}

\section{Performance Evaluation and Simulation Results} \label{sec:results}
\noindent To evaluate the performance of our simulation framework based on the parameters obtained from the previous section (table \ref{tab:idm_parameters}), we designed a dense highway traffic scenario, similar to NGSIM, involving 150 gamma-distributed drivers. The highway scenario was built to be $2 km$ long with vehicles looping back to the start once they have reached the end. This way, we ensured a consistently congested scenario. The simulations have been executed on a PC with an Intel-R CoreTM i7-7700 @ 3.60GHz CPU, 32 GB of RAM, and NVIDIA GeForce GTX 1080 Ti GPU.

We combined the six classes of drivers introduced earlier and analyzed their driving responses in two scenarios: one in which no vehicles have CCW and one in which all vehicles have CCW. To investigate the effectiveness of implemented FCW algorithms in preventing near-crash and crash scenarios, we defined the emergency state (Figure \ref{fig:driving_tasks}) as a driver hard braking reaction when a warning is issued. This emergency state reaction is simulated by pressing the brake pedal all the way down (i.e., the throttle is $-1$) after an initial perception-reaction delay of 1.3s. The maximum braking torque of all vehicles is $1500 Nm$. The work of\cite{Malaterre1998} shows that braking was the primary form of the reaction of many drivers to an FCW alert. In fact, it is the primary objective of any collision avoidance algorithm to determine the right time for braking to avoid a dangerous scenario. The main parameters of the simulation are presented in table \ref{tab:simulation_parameters}. 

\begin{table*}[h]
\normalsize
\caption{Parameters of Intelligent Driver Model for three categories of drivers}
\centering 
\begin{tabular}{lcccc}
\hline \text { Driver Type }  & \text { Aggressive } & \text { Normal } & \text { Conservative } \\
\hline \text { Comfortable Acceleration, }$\left(m / s^{2} \right)$ & $\alpha \in[$1.53,2.75$]$ & $\alpha \in[$1.43,2.59$]$ & $\alpha \in[1.30,2.41]$ \\
\hline \text { Comfortable Deceleration, }$\left(m / s^{2}\right)$ & $\beta \in[1.52,2.73]$ & $\beta \in[1.43,2.59]$ & $\beta \in[1.27,2.41]$ \\
\hline \text { Desired Time Headway, }(s) & $f_{\Gamma}(x \mid \alpha, \beta)<2$ & $f_{\Gamma}(x \mid \alpha, \beta) \leq 3$ & $3 < f_{\Gamma}(x \mid \alpha, \beta)$ \\
\hline \text { Ratio } & 19 \% & 43 \% & 38 \% \\
\hline
\label{tab:idm_parameters}
\end{tabular}
\end{table*}

\begin{table*}
\normalsize
\caption{Simulation Parameters}
\centering 
\begin{tabular}{lc}
\hline \text { Duration, minutes } & 90 \\
\hline \text { Number of Vehicles }  & 150  \\
\text { Exact Number of Drivers } \\
\text { (Aggressive/Normal/Conservative) }  & 27 / 66 / 57 \\
\hline \text { Network's Packet Error Rate (PER)} & 30 \\
\hline \text { Implemented FCW Algorithms }  & \text{CAMP Logistic Regression,} \\
&  \text{NHTSA Driver-Tuned Models} \\
\hline \text { Error-dependent Estimator }  & \text{Constant Acceleration} \\
\hline
\label{tab:simulation_parameters}
\end{tabular}
\end{table*}

In addition to Time Headway ($\tau$), we also use Time-To-Collision (TTC) to evaluate the performance of the two FCW algorithms. TTC is a safety metric that is frequently used in the literature and can be computed using the following formula\cite{Lu2012}:
\begin{equation}
    TTC_i = \frac{x_i - x_{i-1}}{v_i - v_{i-1}}, \ \forall \: v_i > v_{i-1}
\end{equation}
where indices $i$ and $i-1$ represent the host and the leading vehicles, respectively.

We start our first simulation analysis without the warning algorithm enabled. After multiple runs, a factor of $3\%$ distracted drivers resulted in an average of $42$ accidents, $81\%$ of which were resulted from distraction, $17\%$ from aggressive driving characteristics, and the other $2\%$ were caught in a pileup. Of these $42$ counted collisions, $33.3\%$ involved aggressive drivers, $50\%$ normal drivers, and $16.3\%$ conservative drivers.
It is worth mentioning that the vehicles that were involved in an accident continued blocking the road for a period of $10$ to $20s$ before being removed from and respawned back into the simulator at a later time.

\begin{figure}
    \centering
    \includegraphics[width=1\linewidth,trim={5mm 0mm 10mm 0mm},clip]{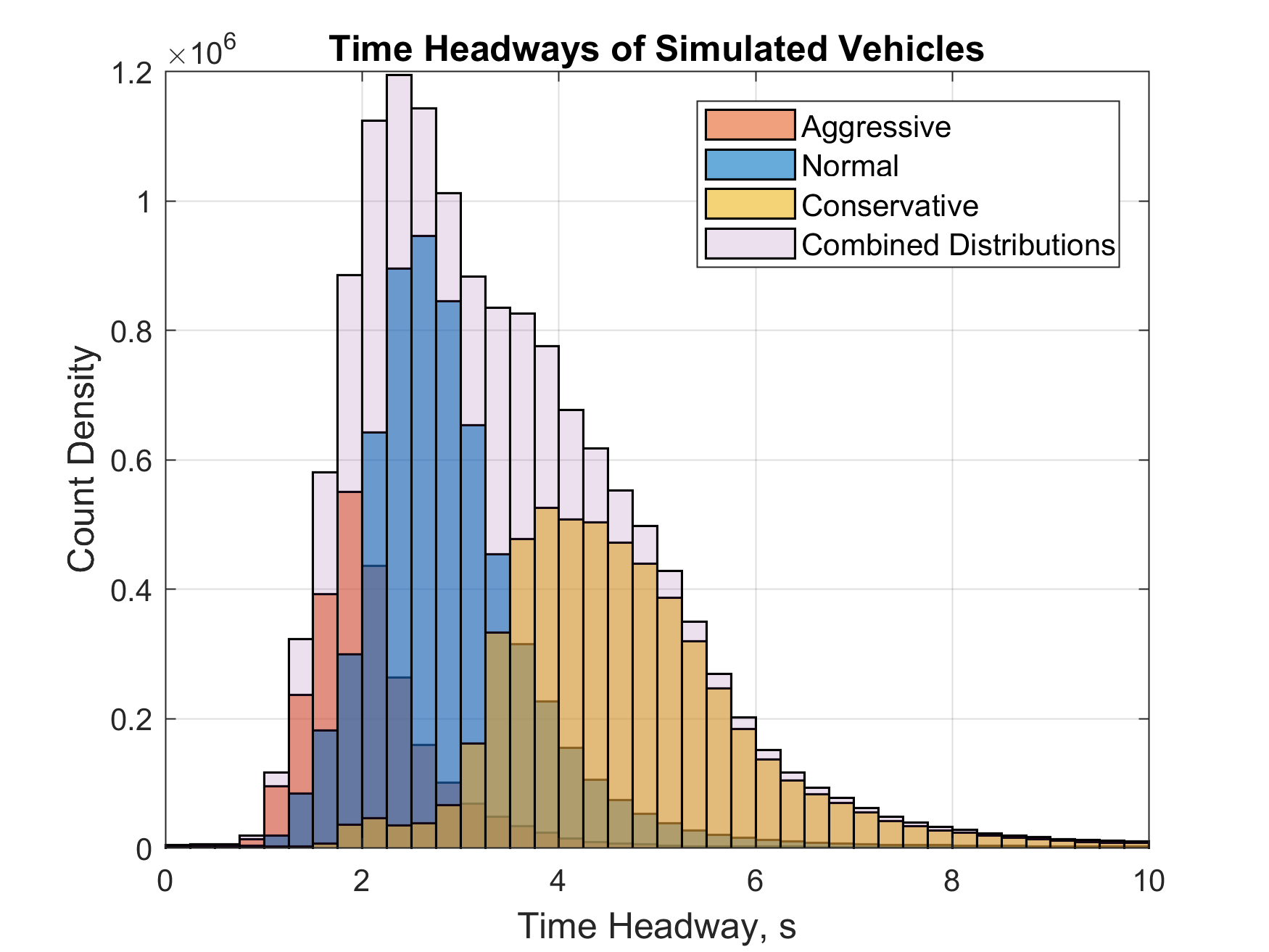}
    \caption{Count density distribution of time headways of simulated vehicles for each class of drivers.}
    \label{fig:count_density_time_headways}
\end{figure}

For a comparison with NGSIM data, the count density distribution of time headways of different classes of drivers is shown below (Figure \ref{fig:count_density_time_headways}). The following results include both cautious and distracted drivers for each behavior characteristic. We can observe that the combined distribution with a mode of $~2.5s$ exhibits similar characteristics as the analysis shown in section \ref{sec:dataset_analysis}.

Next, the same scenario was repeated several times but with the vehicles equipped with FCW systems and the drivers able to react to a warning by hard braking. We run a series of analyses with a communication network packet error rate (PER) of $30\%$. Table \ref{tab:results} shows the collision statistics obtained for each class of driver and the respective FCW algorithm.

\begin{table*}
\caption{Crash and warning statistics for a simulated 90 minutes long scenario}
\begin{adjustbox}{max width=1.0\textwidth}
\centering 
        \begin{tabular}{llcccccccccccc}
        \hline
            \multicolumn{2}{l}{\textbf{Driver Type (Number of Drivers)}} 
            & \multicolumn{3}{c}{\textbf{Aggressive (27)}} & \phantom{abc} 
            & \multicolumn{3}{c}{\textbf{Normal (66)}} & \phantom{abc} 
            & \multicolumn{3}{c}{\textbf{Conservative(57)}} \\  \cmidrule{3-5} \cmidrule{7-9} \cmidrule{11-13}
            
            & Collision Information & Total & Distracted & \begin{tabular}{@{}c@{}}Leader Hard \\ Braking \end{tabular}
            && Total & Distracted & \begin{tabular}{@{}c@{}}Leader Hard \\ Braking \end{tabular} 
            && Total & Distracted & \begin{tabular}{@{}c@{}}Leader Hard \\ Braking \end{tabular} \\ \cmidrule{2-13}
            \parbox[t]{1mm}{\multirow{-3.5}{*}{\rotatebox[origin=r]{90}{Collision Statistics}}}
            & No Warning Algorithm & 14 & 7 & 0 && 21 & 21 & 0 && 7 & 6 & 0 \\
            & CAMP Logistic Regression & 6 & 2 & 4 && 8 & 8 & 1 && 5 & 5 & 0 \\ 
            & NHTSA Early $(0.32g)$ & 6 & 6 & 0 && 5 & 5 & 0 && 0 & 0 & 0 \\ 
            & NHTSA Intermediate $(0.4g)$ & 6 & 3 & 2 && 5 & 5 & 0 && 3 & 3 & 0 \\ 
            & NHTSA Imminent $(0.55g)$ & 11 & 6 & 7 && 7 & 7 & 2 && 1 & 1 & 0 \\
            \hline \hline \\
            & Warning Information & Total & Positive & Ratio 
            && Total & Positive & Ratio  
            && Total & Positive & Ratio \\ \cmidrule{2-13}
            \parbox[t]{1mm}{\multirow{-3}{*}{\rotatebox[origin=c]{90}{Warning Statistics}}} \\
            & CAMP Logistic Regression & 843 & 156 & 18.5\% && 1116 & 411 & 36.8\% && 309 & 38 & 12.3\% \\ 
            & NHTSA Early $(0.32g)$ & 1105 & 117 & 10.5\% && 1063 & 361 & 34.0\% && 193 & 168 & 87.0\% \\ 
            & NHTSA Intermediate $(0.4g)$ & 1290 & 70 & 5.4\% && 958 & 402 & 42.0\% && 208 & 143 & 68.7\% \\ 
            & NHTSA Imminent $(0.55g)$ & 989 & 122 & 12.3\% && 442 & 208 & 47.1\% && 149 & 57 & 38.3\% \\
            
            \hline
        \end{tabular}
        \end{adjustbox}
\label{tab:results}
\end{table*}

While some collisions were inevitable, all the FCW algorithms were effective in significantly reducing the number of collisions. Upon the occurrence of a collision, our algorithms analyzed the scenario to determine who was at fault. The "Total" column for each driver's class in table \ref{tab:results} refers to the total number of crashes that type of driver was at fault. We can see that for all scenarios analyzed, conservative drivers were involved in considerably fewer accidents, whereas aggressive drivers, despite their smaller numbers $(18\%)$, were the most at fault. Surprisingly, all accidents in which a normal or conservative driver was at fault were caused by them being distracted. Hence, this implies that not all the tested FCW algorithms were effective in preventing distracted-related collisions. Recall that all drivers had a $1.3s$ perception-reaction delay. This implies that the implemented algorithms may still fail in complex circumstances, e.g., the vehicle traveling at higher speeds with its driver being distracted while the leader experiences an emergency state and suddenly brakes\cite{9644629}.
Amongst the two FCW algorithms, CAMP Logistic Regression performed worse than all three settings of NHTSA as this algorithm is not a fully-adaptive algorithm with less flexibility compared to the driver-tuned NHTSA ones.

When a driver displays a sudden reaction to a warning by hard braking, it forces its following vehicles to slow down or potentially hard brake too, causing more hazards and near-crash scenarios\cite{9837992}. For the simulation configuration above, we observed that an accident might result from either a distracted driver or a sudden slowdown of a leading vehicle, or a combination of both. Therefore, we consider a warning that successfully averted such a crash scenario into a near-crash as a positive warning. 

Table \ref{tab:results} shows the statistics with regard to the number of warnings generated. For the aggressive driver, we can see that the NHTSA Early and Intermediate had the most warnings generated, though the majority of these were false warnings. This is also true for the other two algorithms which imply that for aggressive driving behavior, none of these algorithms are suitable options. Such a significant fraction of false warnings will result in the frustration of the driver and ultimately in him/her shutting down the system. A similar trend follows for normal drivers whereas the NHTSA with Early alert settings appeared to assist the conservative drivers exceptionally well with $87\%$ positive warnings and zero accidents at fault.

\begin{figure}
    \centering
    \includegraphics[width=1\linewidth,trim={5mm 0mm 10mm 0mm},clip]{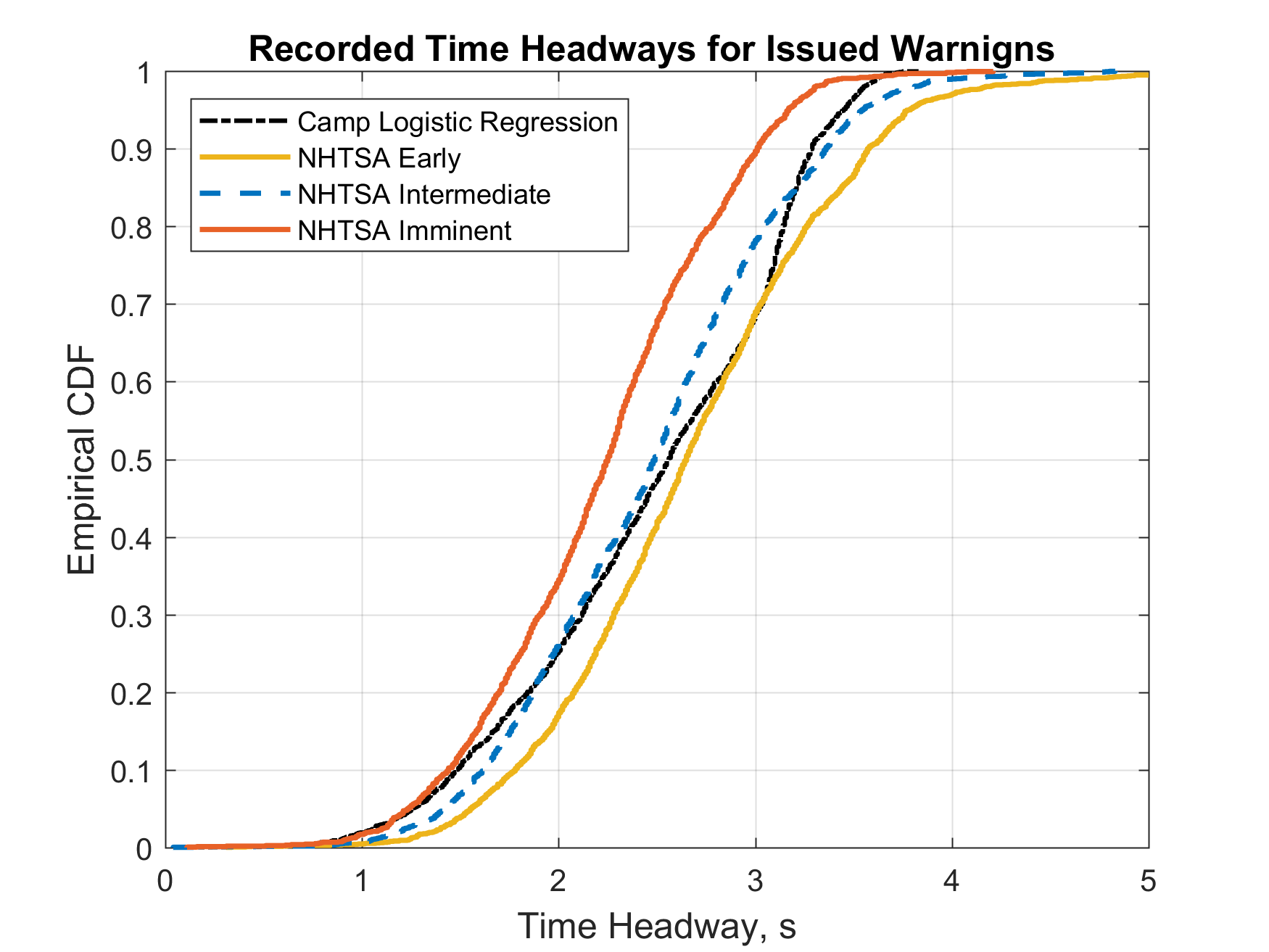}
    \caption{Sensitivity of each forward collision warning algorithm represented through empirical distribution functions of time headway}
    \label{fig:fcw_time_headway_sensitivity}
    \vspace{-0.2 in}
\end{figure}

\begin{figure}
    \centering
    \includegraphics[width=1\linewidth,trim={5mm 0mm 10mm 0mm},clip]{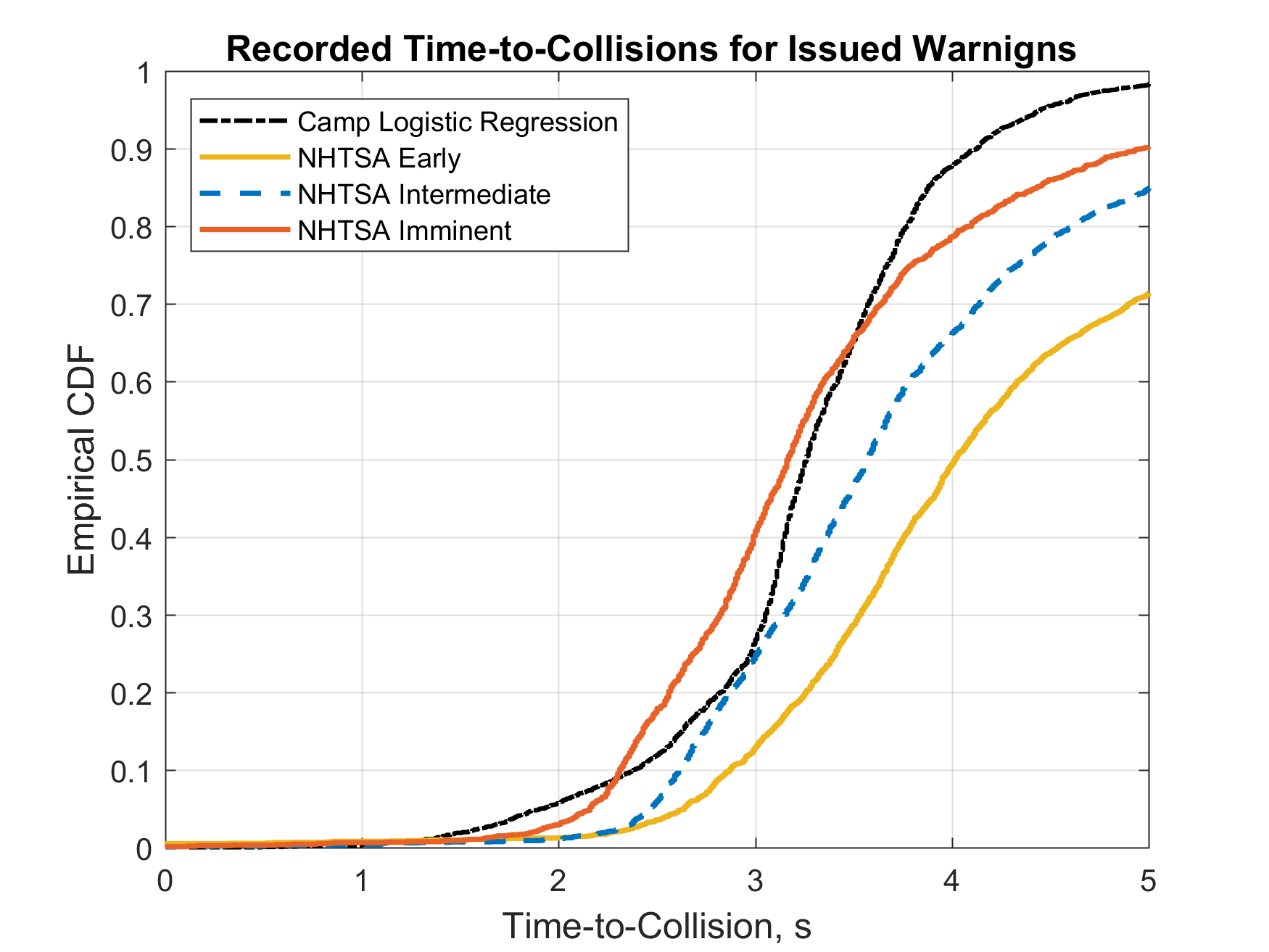}
    \caption{Sensitivity of each forward collision warning algorithm represented through empirical distribution functions of time-to-collision}
    \label{fig:fcw_time_to_collision_sensitivity}
    \vspace{-0.2 in}
\end{figure}

The overall sensitivity of each FCW algorithm is represented as the Probability Distribution Function (PDF) of the TTCs when a warning was issued (Figure \ref{fig:fcw_time_headway_sensitivity}).  While the time headway distributions show similar characteristics and are close to each other, we can see a noticeable difference in the distribution of time-to-collision. Figure \ref{fig:fcw_time_to_collision_sensitivity} explains the poor performance and higher number of accidents for the CAMP Logistic Regression algorithm as the $90^{th}$ percentile of the warnings being generated when host vehicles were less than $4s$ away from colliding with their leading vehicles. We can conclude from this observation that the algorithm was tuned for very aggressive driving characteristics. However, considering the perception-reaction delay of a driver, the timings of these warnings were still late even for an aggressive driver to effectively avoid the collision. This observation implies that it is crucial to consider different driver-specific parameters, especially the perception-reaction time when designing adaptive vehicular active safety systems. Accordingly, this trend is followed by NHTSA Imminent, Intermediate, and with Early having a $90^{th}$ percentile of $8s$. Note that all the time headways and time-to-collisions shown in this paper are calculated via communicated data that are received by host vehicles through BSM messages.

\section{Conclusion and Future Work} \label{sec:conclusion}
\noindent With the rapid growth of advanced driver assistants and automated systems, simulation tools have become an integral part of the development process of these systems. Simulators not only can improve development efficiency, but also can offer opportunities to analyze scenarios that otherwise would have been extremely costly, time-consuming, and on some occasions dangerous to human subjects. In this work, we extended our co-simulation framework \cite{Jami2017} and deployed it on the state-of-the-art game engine technology UE4. With the help of NVIDIA PhysX and the UE4 physics engine, we were able to represent a large-scale traffic scenario with detailed 3D physics and visualization. We described the individual components of our platform in detail and proposed a simple human interpretable extension to the traditional car-following models. The extended model related a human driver to a feedback controller and employed the Intelligent Driver Model to supply a reference acceleration while incorporating a Fuzzy-PD system to compensate for the error. In this extended approach, we integrated the perception-reaction delay time of the driver as an individual component, along with descriptive driving characteristics through the parameters of the Intelligent Driver Model and Fuzzy membership functions. Furthermore, to extract these various driving behaviors, we surveyed different driving characteristics and examined a traffic dataset from a congested scenario. We used the drivers' time headway and acceleration as descriptive features of their driving style. Using this data, we divided the drivers into three categories: aggressive, normal, and conservative. We found that a gamma-distribution function can best describe the dispersal of the drivers within the environment. Finally, we incorporated all the extracted parameters into our simulator and equipped the vehicles with forward collision warning systems. We then analyzed the performance of these systems for each class of driver with a hard-braking response as a driver reaction to a collision warning.
Of the tested safety algorithms, driver-tuned NHTSA with Early warning configuration performed remarkably well for conservative drivers with $87\%$ positive warnings generation and prevented all rear-end collisions. Conversely, none of the four configurations of forward collision warning systems showed adequate performance for aggressive driving characteristics.

Many active safety systems make extensive use of the kinematical parameters of both the host and the leading vehicles. Given the importance of these parameters in the judgment made by safety algorithms, they must be simulated with the highest accuracy possible. Unfortunately, car-following models make significant simplifications to the vehicles' dynamics, to a level that a detailed look at their response curve shows unnatural and abrupt changes in the velocity and acceleration of vehicles, which leads to false warnings even for safe and collision-free scenarios. The unrealistic kinematical response of CFMs especially taints those models that are based on calculating the next speed value. This limitation led to many false warnings in our previous evaluations. Separation of vehicle dynamics from its driver model allowed for a more accurate kinematic response which in turn resulted in better studying of such active safety systems and obtaining test results that were otherwise not producible by traditional simulators.

It is expected that AVs and HVs will share the road. AVs' safety and reliability will be determined by their social awareness and ability to engage in complex social interactions in a socially acceptable manner. However, AVs are still inefficient when it comes to collaborating with HVs, and they struggle to understand and adapt to human behavior, which is especially difficult in mixed autonomy. We intend to continue this work in this direction and improve the simulation environment so that it can incorporate data from real-world traffic and handle more complex interactions between HVs and AVs.

\balance
\bibliography{main.bib}{}
\bibliographystyle{unsrt}
\end{document}